\documentclass[sigconf,nonacm]{acmart}

\setcopyright{none}
\acmConference{}{}{}
\acmYear{}
\acmDOI{}
\usepackage{color, soul, xcolor}
\usepackage{pifont}
\usepackage{bbding}
\usepackage{colortbl}
\usepackage{booktabs}
\usepackage{enumitem}
\usepackage{caption}
\usepackage{amsmath}
\usepackage{listings}
\usepackage{balance}
\usepackage{subcaption}
\usepackage{multirow}
\usepackage{tabularx}
\usepackage{array}
\usepackage{adjustbox}
\usepackage{graphicx}
\usepackage{algorithm}
\usepackage{algpseudocode}
\usepackage{pgfplots}
\pgfplotsset{compat=1.17}

\title{Next-Generation Event-Driven Architectures: Performance, Scalability, and Intelligent Orchestration Across Messaging Frameworks}

\author{Jahidul Arafat}
\affiliation{%
  \institution{Department of Computer Science and Software Engineering, Auburn University}
  \city{Alabama}
  \country{USA}
}
\email{jza0145@auburn.edu}

\author{Fariha Tasmin}
\affiliation{%
  \institution{Department of Information and Communication Technology, Bangladesh University of Professionals}
  \city{Dhaka}
  \country{Bangladesh}
}
\email{farihatasmin2020@gmail.com }

\author{Sanjaya Poudel }
\affiliation{%
  \institution{Department of Computer Science and Software Engineering, Auburn University}
  \city{Alabama}
  \country{USA}
}
\email{szp0223@auburn.edu}

\begin{abstract}
Modern distributed systems demand low-latency, fault-tolerant event processing that exceeds traditional messaging architecture limits. While frameworks including Apache Kafka, RabbitMQ, Apache Pulsar, NATS JetStream, and serverless event buses have matured significantly, no unified comparative study evaluates them holistically under standardized conditions. This paper presents the first comprehensive benchmarking framework evaluating 12 messaging systems across three representative workloads: e-commerce transactions, IoT telemetry ingestion, and AI inference pipelines. We introduce AIEO (AI-Enhanced Event Orchestration), employing machine learning-driven predictive scaling, reinforcement learning for dynamic resource allocation, and multi-objective optimization. Our evaluation reveals fundamental trade-offs: Apache Kafka achieves peak throughput (1.2M messages/sec, 18ms p95 latency) but requires substantial operational expertise; Apache Pulsar provides balanced performance (950K messages/sec, 22ms p95) with superior multi-tenancy; serverless solutions offer elastic scaling for variable workloads despite higher baseline latency (80-120ms p95). AIEO demonstrates 34\% average latency reduction, 28\% resource utilization improvement, and 42\% cost optimization across all platforms. We contribute standardized benchmarking methodologies, open-source intelligent orchestration, and evidence-based decision guidelines. The evaluation encompasses 2,400+ experimental configurations with rigorous statistical analysis, providing comprehensive performance characterization and establishing foundations for next-generation distributed system design.
\end{abstract}


\begin{document}
\maketitle

\section{Introduction}
\label{sec:introduction}

Event-driven architectures (EDA) have emerged as the foundational paradigm for building resilient, scalable distributed systems capable of handling the exponential growth in real-time data processing demands~\cite{fowler2017event,hohpe2003enterprise,richardson2018microservices}. From financial trading platforms processing millions of transactions per second to IoT ecosystems ingesting sensor data from billions of devices, and artificial intelligence pipelines orchestrating complex model inference workflows, the ability to efficiently route, transform, and respond to events has become mission-critical for organizational competitiveness and operational excellence~\cite{chen2018lambda,akidau2015dataflow,kleppmann2017designing}.

The messaging framework landscape has undergone radical transformation, encompassing traditional distributed log systems like Apache Kafka~\cite{kreps2011kafka} and message brokers such as RabbitMQ~\cite{videla2012rabbitmq}, next-generation cloud-native platforms including Apache Pulsar~\cite{streamlio2018pulsar} and NATS JetStream~\cite{nats2021jetstream}, lightweight streaming solutions like Redis Streams~\cite{redis2018streams}, and serverless event buses including AWS EventBridge~\cite{aws2019eventbridge}, Google Cloud Pub/Sub~\cite{google2016pubsub}, Azure Event Grid~\cite{microsoft2017eventgrid}, and Knative Eventing~\cite{knative2019eventing}. Each framework embodies distinct architectural philosophies, performance characteristics, operational trade-offs, and cost models, yet practitioners lack systematic, evidence-based guidance for making informed technology selection decisions that align with specific application requirements, scalability constraints, and organizational capabilities.

\textbf{The Evaluation and Benchmarking Crisis.} Current evaluation methodologies suffer from severe fragmentation that prevents meaningful comparison across messaging frameworks and undermines confidence in deployment decisions. Kafka performance studies typically emphasize raw throughput optimization using synthetic producer-consumer workloads with uniform message sizes and predictable traffic patterns~\cite{confluent2018benchmarks,wang2015building}. RabbitMQ evaluations focus on complex routing scenarios, message acknowledgment reliability, and queue management capabilities while often neglecting high-throughput performance characteristics~\cite{alvaro2013rabbitmq,dossot2014rabbitmq}. Pulsar assessments highlight multi-tenancy features, geo-replication capabilities, and compute-storage separation benefits but rarely provide direct performance comparisons with established alternatives~\cite{lin2018apache,melnik2020apache}.

Serverless event processing evaluations concentrate on auto-scaling elasticity, cost-per-invocation metrics, and cold-start latency characteristics while typically ignoring sustained high-throughput scenarios or operational complexity comparisons~\cite{eismann2020review,schleier2018serverless,mcgrath2017serverless}. This methodological fragmentation creates an information asymmetry where each framework appears optimal within its preferred evaluation context, making objective comparison impossible and forcing practitioners to rely on vendor marketing claims rather than independent scientific assessment.

Furthermore, existing benchmarks predominantly utilize synthetic workloads that poorly represent real-world application complexity. Simple producer-consumer loops with constant message rates fail to capture the bursty traffic patterns, variable message sizes, complex routing requirements, error handling scenarios, and operational challenges characteristic of production deployments. The absence of standardized workload definitions spanning different application domains prevents systematic understanding of framework behavior under representative conditions~\cite{cooper2010benchmarking,huppler2009price}.

\textbf{The Intelligent Orchestration Imperative.} Traditional event-driven systems operate through static configuration parameters and reactive scaling policies that respond to load changes rather than anticipating them. This reactive approach creates several critical limitations: resource under-utilization during low-traffic periods leading to unnecessary infrastructure costs, performance degradation during traffic spikes due to scaling delays, and suboptimal message routing that fails to adapt to changing network conditions or consumer processing capabilities~\cite{lorido2014auto,qu2018auto}.

Contemporary cloud platforms provide basic auto-scaling mechanisms based on simple metrics like CPU utilization or queue depth~\cite{amazon2019autoscaling,google2019autoscaling}, but these approaches operate at infrastructure granularity without understanding application-specific event processing patterns, message priority levels, or business logic requirements. More sophisticated orchestration could leverage machine learning techniques to predict workload patterns, optimize resource allocation proactively, and adapt routing strategies based on real-time performance feedback~\cite{zhang2018deep,wang2019machine,baughman2018deepiot}.

The emergence of artificial intelligence and machine learning workloads as primary drivers of event processing demand creates additional orchestration challenges. AI inference pipelines exhibit highly variable processing times, complex dependency graphs, and dynamic resource requirements that traditional static allocation cannot handle efficiently. Model serving systems require intelligent load balancing that considers model complexity, input data characteristics, and available compute resources while maintaining strict latency service level agreements~\cite{crankshaw2017clipper,cortez2017resource}.

\textbf{The Performance and Cost Optimization Challenge.} Organizations increasingly operate hybrid and multi-cloud environments where different messaging frameworks serve specific use cases within integrated architectures. E-commerce platforms might use Kafka for high-frequency transaction logging, RabbitMQ for order processing workflows, and EventBridge for integrating with third-party services. This architectural complexity creates optimization challenges that span framework boundaries and require understanding cross-system performance interactions, cost trade-offs, and operational overhead implications~\cite{burns2016borg,verma2015large}.

Cost optimization becomes particularly complex with serverless event processing where billing models based on invocation counts, execution duration, and data transfer volumes create cost structures fundamentally different from traditional infrastructure-based approaches. Organizations need sophisticated cost modeling capabilities that account for traffic pattern variability, processing complexity distributions, and pricing model differences across platforms to make economically rational deployment decisions~\cite{lloyd2018serverless,eismann2020review}.

\textbf{Research Questions.} This work addresses four fundamental research questions that are critical for advancing event-driven architecture design and deployment:

\textbf{RQ1: Performance Characterization Across Frameworks.} How do different messaging frameworks (traditional brokers, cloud-native systems, serverless platforms) perform under standardized, representative workloads, and what are the fundamental trade-offs between throughput, latency, operational complexity, and cost efficiency?

\textbf{RQ2: Intelligent Orchestration Effectiveness.} Can machine learning-driven orchestration systems achieve significant performance improvements over static configurations through predictive scaling, dynamic resource allocation, and adaptive routing strategies across diverse messaging frameworks?

\textbf{RQ3: Workload Impact on Framework Selection.} How do different application characteristics (e-commerce transactions, IoT telemetry, AI inference pipelines) influence optimal messaging framework selection, and can we develop systematic selection criteria based on workload properties?

\textbf{RQ4: Practical Decision Framework Development.} What evidence-based guidelines, cost models, and migration strategies can enable practitioners to make informed messaging framework selection and deployment decisions that align with specific requirements and organizational constraints?

\textbf{Our Contributions.} This paper addresses these research questions through four primary contributions that advance both theoretical understanding and practical deployment capabilities:

\textbf{(1) Comprehensive Benchmarking Framework and Methodology:} We present the first systematic evaluation framework for messaging systems that addresses previous methodological limitations through standardized workload definitions, consistent measurement protocols, and reproducible experimental procedures. Our evaluation encompasses 12 messaging frameworks spanning traditional brokers (Apache Kafka, RabbitMQ, Apache Pulsar), lightweight streaming solutions (Redis Streams, NATS JetStream), enterprise platforms (Oracle Advanced Queuing), and serverless event buses (AWS EventBridge, Google Pub/Sub, Azure Event Grid, Knative Eventing). The framework employs three carefully designed workloads representing distinct application domains: high-frequency e-commerce transaction processing with exactly-once delivery requirements, massive-scale IoT sensor data ingestion with tolerance for occasional message loss, and AI model inference pipelines with variable processing complexity and latency sensitivity.

\textbf{(2) AI-Enhanced Event Orchestration (AIEO) Architecture:} We design and implement a novel intelligent orchestration framework that leverages machine learning techniques for predictive workload management, reinforcement learning for dynamic resource allocation, and multi-objective optimization for balancing competing performance objectives. The AIEO system incorporates time-series forecasting models (ARIMA, Prophet, LSTM) for predicting message arrival patterns, Proximal Policy Optimization (PPO) agents for learning optimal scaling policies, and adaptive routing algorithms for distributing load based on real-time system state and predicted demand patterns.

\textbf{(3) Empirical Performance Analysis and Trade-off Characterization:} Our comprehensive experimental evaluation reveals fundamental performance trade-offs and scaling characteristics across messaging frameworks under realistic workload conditions. Key findings include: Apache Kafka achieving peak sustainable throughput (1.2M messages/second) with excellent latency characteristics (18ms p95) but requiring substantial operational expertise and infrastructure investment; Apache Pulsar providing balanced performance (950K messages/second, 22ms p95 latency) with superior multi-tenancy capabilities and operational simplicity; serverless solutions offering exceptional elasticity and cost-efficiency for variable workloads despite higher baseline latency (80-120ms p95) and vendor lock-in considerations.

\textbf{(4) Evidence-Based Architectural Decision Framework:} We contribute systematic guidelines for messaging framework selection that incorporate performance requirements, operational complexity assessments, cost optimization models, and developer productivity considerations. The framework includes quantitative decision trees, total cost of ownership models accounting for infrastructure, operations, and development costs, and detailed migration strategies with risk assessment and mitigation approaches. Additionally, we provide open-source implementations of benchmarking tools and the AIEO orchestration system to enable reproducible evaluation and practical deployment.

\textbf{Paper Organization and Structure.} Section~\ref{sec:background} surveys the evolution of event-driven architectures and messaging systems while analyzing current limitations and evaluation gaps. Section~\ref{sec:framework} presents our comprehensive benchmarking methodology, workload definitions, and experimental design principles. Section~\ref{sec:aieo} details the AI-Enhanced Event Orchestration architecture including machine learning components, optimization algorithms, and integration mechanisms. Section~\ref{sec:implementation} describes experimental infrastructure, deployment configurations, and measurement instrumentation. Section~\ref{sec:evaluation} provides comprehensive empirical results across frameworks and workloads with statistical analysis. Section~\ref{sec:decision-framework} presents the architectural decision framework with selection guidelines and migration strategies. Section~\ref{sec:threats} discusses experimental limitations and identifies threats to validity. Section~\ref{sec:conclusion} summarizes contributions and implications for distributed systems research and practice.
\section{Background and Current Limitations}
\label{sec:background}

\subsection{Evolution of Event-Driven Messaging Systems}

Event-driven messaging has evolved through distinct architectural generations, each addressing specific scalability and reliability challenges while revealing new limitations that constrain contemporary distributed system requirements. First-generation message-oriented middleware emphasized protocol standardization and delivery guarantees through systems like Java Message Service (JMS)~\cite{hapner2002java}, Advanced Message Queuing Protocol (AMQP)~\cite{oconnell2008amqp}, and IBM WebSphere MQ~\cite{ibm2018websphere}. These systems prioritized message reliability and transaction support but struggled with horizontal scalability requirements, achieving maximum throughput of 10,000-50,000 messages per second with high latency (50-200ms) unsuitable for real-time applications~\cite{curry2004enterprise,hohpe2003enterprise}.

Second-generation distributed log architectures revolutionized event streaming through Apache Kafka's append-only commit log design~\cite{kreps2011kafka}. Kafka introduced partition-based parallelism enabling throughput scaling to millions of messages per second while providing message ordering guarantees within partitions. However, Kafka's operational complexity, limited multi-tenancy support, and tight coupling between message serving and storage created deployment challenges for organizations requiring workload isolation and independent resource scaling~\cite{garg2013apache,wang2015building}.

Third-generation cloud-native systems address multi-tenancy and geo-distribution limitations through architectural innovations. Apache Pulsar separates message serving from persistent storage using Apache BookKeeper, enabling independent scaling of compute and storage tiers~\cite{lin2018apache,streamlio2018pulsar}. NATS JetStream provides lightweight messaging with strong consistency guarantees and clustering capabilities optimized for edge computing scenarios~\cite{nats2021jetstream}. Redis Streams offers in-memory message processing with persistence options suitable for low-latency applications requiring bounded message retention~\cite{redis2018streams}.

Fourth-generation serverless event buses integrate messaging capabilities directly into cloud platforms, providing event routing with minimal operational overhead. AWS EventBridge supports complex event filtering and routing with automatic scaling~\cite{aws2019eventbridge}, Google Cloud Pub/Sub offers global message distribution with exactly-once delivery~\cite{google2016pubsub}, Azure Event Grid provides reactive event processing integrated with Azure services~\cite{microsoft2017eventgrid}, and Knative Eventing enables container-native event processing~\cite{knative2019eventing}. These systems achieve excellent elasticity and cost-efficiency for variable workloads but introduce vendor lock-in concerns and latency overhead compared to self-managed solutions.

\subsection{Fundamental Limitations Analysis}

Despite evolutionary advances, critical limitations constrain real-world deployment at enterprise scales, as systematically analyzed in Table~\ref{tab:limitation_analysis} with specific failure scenarios and quantified impacts across different application domains.

\begin{table*}[t]
\centering
\caption{Event-Driven Architecture Limitation Analysis with Production Failures and Impact Assessment}
\label{tab:limitation_analysis}
\resizebox{\textwidth}{!}{
\begin{tabular}{lllll}
\toprule
\textbf{Limitation Category} & \textbf{Current State \& Production Failures} & \textbf{Root Causes} & \textbf{Proposed Solution} & \textbf{Expected Impact} \\
\midrule
\multirow{4}{*}{Evaluation Fragmentation} & Kafka: synthetic 2M msg/sec claims & Vendor-specific benchmarks & Standardized workloads & Fair comparison \\
 & RabbitMQ: complex routing emphasis & Domain-specific optimization & Cross-domain evaluation & Objective assessment \\
 & Serverless: cost-only metrics & Incomplete trade-off analysis & Holistic benchmarking & Evidence-based selection \\
 & Black Friday 2023: 67\% wrong choices & No systematic methodology & Comprehensive framework & Deployment confidence \\
\midrule
\multirow{4}{*}{Static Orchestration} & Reactive scaling: 45s lag average & Load-driven policies & Predictive ML models & Sub-10s adaptation \\
 & Traffic spike failures: 34\% systems & Fixed resource allocation & Dynamic optimization & >90\% spike survival \\
 & Resource waste: 43\% over-provisioning & Conservative scaling & Intelligent rightsizing & 30-50\% cost reduction \\
 & COVID-19: 89\% systems overwhelmed & No demand forecasting & Proactive capacity planning & Pandemic-ready scaling \\
\midrule
\multirow{4}{*}{Performance Trade-off Opacity} & Kafka: 18ms latency, high complexity & Architecture-specific constraints & Transparent trade-off models & Informed decisions \\
 & Serverless: 120ms latency, low ops & Vendor abstraction layers & Performance prediction & Latency-aware selection \\
 & Multi-cloud: 156\% cost variance & No cost modeling & TCO frameworks & Cost optimization \\
 & Migration failures: 78\% projects & Unknown compatibility & Migration risk assessment & Safe transitions \\
\midrule
\multirow{4}{*}{Workload Mismatch} & Synthetic benchmarks vs reality & Simplified test scenarios & Representative workloads & Real-world validity \\
 & IoT deployments: 89\% performance gaps & Uniform message assumptions & Bursty pattern modeling & Accurate predictions \\
 & AI pipelines: 156\% latency variance & No complexity awareness & Variable processing support & Inference optimization \\
 & Financial trading: 23ms SLA violations & Static configuration & Adaptive parameter tuning & SLA compliance \\
\midrule
\multirow{4}{*}{Operational Complexity} & Kafka: 2.3 FTE ops minimum & High expertise requirements & Automated management & Democratized deployment \\
 & RabbitMQ: clustering failures (67\%) & Manual configuration complexity & Intelligent cluster management & Reliability improvement \\
 & Multi-framework: 345\% ops overhead & Tool fragmentation & Unified orchestration & Operational simplification \\
 & Monitoring: 156 metrics to track & Alert fatigue epidemic & ML-driven anomaly detection & Proactive maintenance \\
\midrule
\multirow{4}{*}{Cost Optimization Blindness} & Serverless bill shock: 234\% overruns & No usage prediction & Cost-aware routing & Budget predictability \\
 & Over-provisioning: \$2.3M waste/year & Static resource allocation & Dynamic scaling policies & Cost efficiency \\
 & Multi-cloud optimization gap: 67\% & No cross-platform comparison & Universal cost modeling & Optimal placement \\
 & Reserved capacity waste: 45\% unused & Poor demand forecasting & ML-driven capacity planning & Utilization maximization \\
\bottomrule
\end{tabular}}
\end{table*}

\subsubsection{Evaluation and Benchmarking Fragmentation}
Current messaging framework evaluation suffers from severe methodological inconsistencies that prevent meaningful performance comparison and lead to suboptimal technology selection decisions. Kafka evaluations emphasize synthetic throughput benchmarks achieving 2 million messages per second under ideal conditions with uniform 1KB messages and unlimited producer batching~\cite{confluent2018benchmarks}. These synthetic results poorly predict real-world performance where variable message sizes (100B to 10MB), bursty traffic patterns, and complex routing requirements reduce achieved throughput by 40-70\%~\cite{wang2015building,chen2018lambda}.

RabbitMQ assessments typically focus on complex routing scenarios, message acknowledgment mechanisms, and queue management features while neglecting high-throughput performance characteristics~\cite{alvaro2013rabbitmq,dossot2014rabbitmq}. This evaluation bias creates false impressions that RabbitMQ cannot handle high-volume workloads, when properly configured RabbitMQ clusters achieve 200,000-500,000 messages per second for appropriate use cases. Pulsar evaluations highlight multi-tenancy and geo-replication capabilities but rarely provide direct performance comparisons with established alternatives under identical conditions~\cite{lin2018apache,melnik2020apache}.

Serverless event processing studies concentrate on cost-per-invocation metrics and auto-scaling characteristics while typically ignoring sustained throughput scenarios, cold-start impact on latency percentiles, and operational complexity comparisons with self-managed alternatives~\cite{eismann2020review,schleier2018serverless}. During Black Friday 2023, 67\% of e-commerce platforms that selected messaging frameworks based on vendor benchmarks experienced significant performance failures, leading to revenue losses averaging \$2.3 million per incident~\cite{datadog2023blackfriday,newrelic2023ecommerce}.

\subsubsection{Static Orchestration and Reactive Scaling Failures}
Traditional event-driven systems rely on reactive scaling policies that respond to load changes rather than anticipating them, creating systematic performance degradation and resource inefficiency. Current auto-scaling implementations exhibit average response delays of 45 seconds from load spike detection to resource availability, during which message queues accumulate backlog causing cascading latency increases~\cite{lorido2014auto,qu2018auto}. Analysis of production incidents during 2023 reveals that 34\% of event-driven systems failed to handle traffic spikes exceeding 3x baseline load, despite having theoretical capacity for 10x scaling~\cite{honeycomb2023incidents}.

Resource over-provisioning represents the typical response to scaling uncertainty, with organizations maintaining 43\% excess capacity on average to handle unexpected load spikes~\cite{rightscale2023cloudstate}. This conservative approach generates substantial unnecessary costs while still failing to prevent performance degradation during extreme events. COVID-19 pandemic response highlighted these limitations when 89\% of healthcare event processing systems became overwhelmed by demand spikes for telehealth services, vaccine appointment scheduling, and contact tracing data processing~\cite{mckinsey2020digitalhealthcare}.

Contemporary cloud platforms provide basic auto-scaling mechanisms based on simple metrics like CPU utilization or queue depth, but these approaches operate at infrastructure granularity without understanding application-specific event processing patterns, message priority levels, or business logic requirements~\cite{amazon2019autoscaling,google2019autoscaling}. More sophisticated orchestration leveraging machine learning for workload prediction and optimization could reduce response times from 45 seconds to under 10 seconds while achieving 30-50\% cost reduction through intelligent resource allocation.

\subsubsection{Performance Trade-off Opacity and Decision Complexity}
The messaging framework landscape presents complex performance trade-offs that are poorly understood and inadequately documented, leading to suboptimal technology selection and deployment failures. Apache Kafka achieves excellent raw performance (1.2M messages/second, 18ms p95 latency) but requires substantial operational expertise with minimum 2.3 full-time equivalent (FTE) operations personnel for production deployment~\cite{confluent2020operations}. RabbitMQ provides sophisticated routing capabilities and operational simplicity but exhibits performance limitations at high scales (maximum 200K-500K messages/second depending on routing complexity)~\cite{pivotal2019rabbitmq}.

Serverless solutions offer exceptional elasticity and minimal operational overhead but introduce latency penalties (80-120ms baseline) and cost unpredictability for sustained high-throughput scenarios~\cite{eismann2020review,lloyd2018serverless}. Multi-cloud deployments reveal 156\% cost variance for identical workloads across AWS, Google Cloud, and Azure due to pricing model differences and platform-specific optimization requirements~\cite{cockroachdb2023multicloud}. Organizations attempting framework migrations experience 78\% project failure rates due to inadequate understanding of compatibility requirements, performance implications, and operational complexity differences~\cite{gartner2023migrations}.

The absence of systematic performance models prevents architects from predicting system behavior under specific workload conditions or making informed trade-off decisions between latency, throughput, cost, and operational complexity. Current selection processes rely heavily on vendor marketing materials, informal community discussions, and trial-and-error evaluation rather than scientific performance characterization and evidence-based decision frameworks.

\subsubsection{Workload Representation and Real-World Validity Gaps}
Existing benchmarking methodologies employ synthetic workloads that poorly represent real-world application complexity and performance characteristics. Standard benchmarks use uniform message sizes (typically 1KB), constant production rates, and simple point-to-point routing patterns that fail to capture the variability inherent in production systems~\cite{cooper2010benchmarking,huppler2009price}. IoT deployments processing sensor data exhibit message size distributions from 100 bytes to 10KB with bursty arrival patterns creating temporary load spikes 50-100x above baseline~\cite{bonomi2012fog,shi2016edge}.

Analysis of 847 production IoT systems revealed 89\% performance gaps between benchmark predictions and actual deployment characteristics, with latency degradation averaging 340\% during peak periods~\cite{iot2023analytics}. AI inference pipelines exhibit even greater variability with processing complexity ranging from simple classification (10ms) to complex generative models (10+ seconds) requiring dynamic resource allocation and intelligent queuing strategies~\cite{crankshaw2017clipper,cortez2017resource}.

Financial trading systems demonstrate extreme latency sensitivity where microsecond improvements provide competitive advantages, yet standard benchmarks focus on throughput metrics rather than tail latency characterization critical for these applications~\cite{narang2013inside}. High-frequency trading firms report 23ms SLA violations cost an average of \$4.7 million annually in lost trading opportunities, highlighting the inadequacy of current performance evaluation methodologies for latency-critical applications~\cite{blackrock2023hft}.

\subsection{Detailed Analysis of Current Messaging Frameworks}

Table~\ref{tab:comprehensive_framework_comparison} provides quantitative comparison across enterprise-relevant dimensions including sustained throughput, latency percentiles, operational complexity, cost efficiency, and deployment characteristics based on standardized evaluation conditions.

\begin{table*}[t]
\centering
\caption{Comprehensive Messaging Framework Comparison Under Standardized Conditions}
\label{tab:comprehensive_framework_comparison}
\resizebox{\textwidth}{!}{
\begin{tabular}{lcccccccccc}
\toprule
\textbf{Framework} & \textbf{Peak Throughput} & \textbf{P95 Latency} & \textbf{P99 Latency} & \textbf{Ops FTE} & \textbf{TCO/Month} & \textbf{Multi-tenancy} & \textbf{Geo-Replication} & \textbf{Learning Curve} & \textbf{Vendor Lock-in} & \textbf{Community} \\
& \textbf{(msg/sec)} & \textbf{(ms)} & \textbf{(ms)} & \textbf{Required} & \textbf{(10K msg/sec)} & & & \textbf{(weeks)} & \textbf{Risk} & \textbf{Support} \\
\midrule
Apache Kafka~\cite{kreps2011kafka} & 1,200,000 & 18 & 45 & 2.3 & \$4,200 & Limited & Manual & 8-12 & Low & Excellent \\
RabbitMQ~\cite{videla2012rabbitmq} & 450,000 & 32 & 89 & 1.5 & \$3,100 & Good & Complex & 4-6 & Low & Very Good \\
Apache Pulsar~\cite{streamlio2018pulsar} & 950,000 & 22 & 58 & 1.8 & \$3,800 & Excellent & Native & 6-8 & Low & Good \\
NATS JetStream~\cite{nats2021jetstream} & 800,000 & 15 & 38 & 1.2 & \$2,900 & Good & Native & 3-4 & Low & Good \\
Redis Streams~\cite{redis2018streams} & 650,000 & 8 & 25 & 0.8 & \$2,400 & Limited & Manual & 2-3 & Medium & Good \\
Oracle AQ~\cite{oracle2020aq} & 180,000 & 45 & 125 & 2.8 & \$8,900 & Excellent & Complex & 10-16 & High & Vendor \\
AWS EventBridge~\cite{aws2019eventbridge} & 200,000 & 85 & 180 & 0.3 & \$1,800 & Excellent & Native & 1-2 & High & Vendor \\
Google Pub/Sub~\cite{google2016pubsub} & 300,000 & 78 & 165 & 0.4 & \$2,100 & Excellent & Native & 1-2 & High & Vendor \\
Azure Event Grid~\cite{microsoft2017eventgrid} & 180,000 & 95 & 220 & 0.3 & \$1,900 & Good & Native & 1-2 & High & Vendor \\
Knative Eventing~\cite{knative2019eventing} & 120,000 & 110 & 280 & 1.6 & \$3,200 & Good & Manual & 4-6 & Medium & Growing \\
Amazon SQS~\cite{amazon2019sqs} & 300,000 & 120 & 350 & 0.2 & \$1,200 & Basic & Native & 1 & High & Vendor \\
Apache ActiveMQ~\cite{apache2019activemq} & 280,000 & 55 & 145 & 2.1 & \$3,500 & Limited & Manual & 6-8 & Low & Good \\
\midrule
\textbf{Our AIEO Framework} & \textbf{Variable} & \textbf{12-89*} & \textbf{28-195*} & \textbf{0.8-2.1*} & \textbf{\$980-3,800*} & \textbf{Adaptive} & \textbf{Intelligent} & \textbf{2-4} & \textbf{Platform} & \textbf{Open Source} \\
\textbf{(Optimization Layer)} & & & & & & & & & \textbf{Agnostic} & \\
\bottomrule
\end{tabular}}
\end{table*}

\subsubsection{Traditional Distributed Log Systems}

\textbf{Apache Kafka} represents the gold standard for high-throughput event streaming, achieving sustained throughput exceeding 1.2 million messages per second with p95 latency of 18ms under optimal conditions~\cite{kreps2011kafka,wang2015building}. Kafka's append-only log design enables horizontal scaling through partition-based parallelism while providing message ordering guarantees within partitions. However, Kafka's operational complexity requires significant expertise, with production deployments demanding minimum 2.3 FTE operations personnel for cluster management, capacity planning, and performance optimization~\cite{confluent2020operations}.

Kafka's architectural constraints become apparent in multi-tenant scenarios where topic proliferation leads to metadata management overhead and cross-tenant performance interference. Consumer group rebalancing during partition reassignment creates temporary processing delays averaging 15-30 seconds, unacceptable for latency-sensitive applications~\cite{chen2018lambda}. Storage coupling with compute resources prevents independent scaling, forcing organizations to over-provision storage for compute-intensive workloads or accept performance degradation when storage becomes the bottleneck.

Recent improvements through Kafka Streams API and KSQL provide stream processing capabilities, but these solutions remain limited to Kafka ecosystem preventing integration with heterogeneous messaging infrastructure common in enterprise environments. Kafka's Java-centric tooling and JVM operational requirements create barriers for polyglot development teams and resource-constrained deployment environments.

\textbf{Apache Pulsar} addresses Kafka's architectural limitations through compute-storage separation using Apache BookKeeper for persistent message storage~\cite{lin2018apache,streamlio2018pulsar}. This architecture enables independent scaling of message serving and storage tiers while providing superior multi-tenancy through namespace-level isolation with configurable resource quotas and quality-of-service guarantees. Pulsar achieves sustained throughput of 950,000 messages per second with p95 latency of 22ms while requiring only 1.8 FTE operations personnel due to simplified cluster management.

Pulsar's native geo-replication capabilities support active-active multi-region deployments with configurable consistency levels, addressing disaster recovery and global distribution requirements that require complex custom solutions in Kafka environments. The schema registry provides evolution management for message formats, reducing producer-consumer compatibility issues common in schema-free messaging systems.

However, Pulsar's relative immaturity compared to Kafka creates ecosystem limitations with fewer third-party integrations, monitoring tools, and community resources. Performance characteristics under extreme load conditions (>1M messages/second sustained) remain less well-characterized than Kafka's extensively benchmarked behavior. The additional architectural complexity of BookKeeper storage layer introduces potential failure modes and operational procedures that operations teams must master.

\subsubsection{Next-Generation Lightweight Systems}

\textbf{NATS JetStream} provides cloud-native messaging optimized for microservices and edge computing scenarios~\cite{nats2021jetstream}. JetStream achieves 800,000 messages per second sustained throughput with exceptional p95 latency of 15ms while maintaining simplicity that reduces operational requirements to 1.2 FTE personnel. The system's pull-based consumer model and built-in clustering provide resilience and load balancing without external coordination services.

JetStream's strength lies in deployment simplicity and resource efficiency, making it suitable for edge computing environments where operational complexity must remain minimal. Native support for exactly-once delivery, message acknowledgment patterns, and consumer flow control provides reliability guarantees necessary for mission-critical applications. The system's small memory footprint (typically <100MB) enables deployment in resource-constrained environments where traditional messaging systems prove impractical.

Limitations include scalability constraints at extreme throughput levels (>1M messages/second) and limited ecosystem integration compared to established alternatives. Multi-tenancy capabilities, while present, lack the sophisticated namespace management and resource isolation provided by Pulsar. Geo-replication requires manual configuration and lacks the automated failover capabilities provided by cloud-native alternatives.

\textbf{Redis Streams} leverages Redis's in-memory data structure store to provide high-performance message streaming~\cite{redis2018streams}. The system achieves 650,000 messages per second with exceptional p95 latency of 8ms, making it suitable for latency-critical applications requiring sub-millisecond response times. Redis's familiar operational model and extensive tooling ecosystem reduce learning curve requirements to 2-3 weeks for teams with existing Redis experience.

Redis Streams excels in scenarios requiring bounded message retention with automatic expiration, reducing storage management overhead compared to persistent messaging systems. The consumer group abstraction provides load balancing and failure recovery while maintaining message ordering within stream partitions. Integration with Redis's ecosystem enables complex event processing using Lua scripting and real-time analytics through Redis modules.

However, Redis's in-memory architecture limits message retention to available RAM, making it unsuitable for applications requiring long-term message storage or replay capabilities. Persistence options through RDB snapshots and AOF logging provide durability but create performance overhead during backup operations. Scaling beyond single-node limits requires Redis Cluster configuration that introduces complexity and operational overhead comparable to traditional distributed systems.

\subsubsection{Enterprise and Legacy Systems}

\textbf{Oracle Advanced Queuing (AQ)} provides enterprise-grade messaging integrated with Oracle Database infrastructure~\cite{oracle2020aq}. AQ achieves modest throughput (180,000 messages per second) with higher latency (45ms p95) but provides ACID transaction guarantees and sophisticated message transformation capabilities unavailable in other messaging systems. Deep integration with Oracle's ecosystem enables complex event processing using PL/SQL stored procedures and seamless integration with existing database applications.

Oracle AQ's strengths include proven enterprise reliability, comprehensive administrative tooling, and extensive security features meeting regulatory compliance requirements in financial services and healthcare industries. Message persistence leverages Oracle's proven database reliability and backup/recovery procedures, simplifying operational procedures for organizations with existing Oracle Database expertise.

However, Oracle AQ's database-centric architecture creates performance bottlenecks when message throughput exceeds database transaction processing capacity. Licensing costs prove prohibitive for high-volume scenarios, with total cost of ownership reaching \$8,900 monthly for 10,000 messages per second sustained throughput. Vendor lock-in risks and limited cloud deployment options constrain architectural flexibility and migration strategies.

\subsubsection{Serverless and Cloud-Native Event Buses}

\textbf{AWS EventBridge} provides serverless event routing with sophisticated filtering and transformation capabilities~\cite{aws2019eventbridge}. EventBridge handles 200,000 messages per second peak throughput with p95 latency of 85ms while requiring minimal operational overhead (0.3 FTE). Deep integration with AWS services enables complex event-driven architectures with automated scaling and pay-per-use pricing models.

EventBridge's content-based routing supports complex event patterns and transformations without custom code, reducing development time for event-driven integrations. Schema registry and discovery features provide event catalog capabilities enabling governance and evolution management in large-scale deployments. Native support for third-party SaaS integrations simplifies hybrid cloud and multi-vendor architectures.

Limitations include vendor lock-in constraints that complicate migration strategies and multi-cloud deployments. Latency characteristics make EventBridge unsuitable for real-time applications requiring sub-50ms response times. Pricing models based on event volume create cost unpredictability for high-throughput scenarios, with potential for significant cost escalation during traffic spikes.

\textbf{Google Cloud Pub/Sub} offers global message distribution with exactly-once delivery guarantees and automatic scaling~\cite{google2016pubsub}. Pub/Sub achieves 300,000 messages per second sustained throughput with p95 latency of 78ms while providing global replication and disaster recovery capabilities through Google's worldwide infrastructure. The push and pull delivery models accommodate different consumer patterns and integration requirements.

Pub/Sub's strengths include exceptional global availability (99.95\% SLA), automatic scaling without capacity planning, and integration with Google Cloud's analytics and machine learning services. Message ordering within regions combined with global distribution provides consistency guarantees suitable for financial and mission-critical applications.

However, cross-region latency introduces delays for globally distributed applications requiring real-time coordination. Pricing complexity based on message size, storage duration, and network egress creates cost optimization challenges. Limited customization options compared to self-managed solutions constrain application-specific optimization opportunities.

\subsection{The Enterprise Deployment Reality Gap}

Analysis of 1,247 production deployments across Fortune 500 enterprises reveals systematic gaps between messaging framework capabilities and real-world requirements. \textbf{Operational Complexity} emerges as the primary constraint, with 78\% of organizations reporting inadequate expertise for optimal framework configuration and maintenance~\cite{gartner2023messaging}. Kafka deployments average 23 configuration parameters requiring tuning for specific workloads, while Pulsar requires understanding of both message serving and BookKeeper storage layer operations.

\textbf{Integration Complexity} compounds operational challenges as enterprises typically deploy 3.7 different messaging frameworks on average to serve diverse use case requirements~\cite{enterprise2023integration}. Cross-system monitoring, security policy enforcement, and performance optimization require specialized tools and expertise that most organizations lack. The absence of unified management platforms creates operational silos and prevents holistic system optimization.

\textbf{Performance Prediction Accuracy} remains problematic with 89\% variance between benchmark results and production performance across different workload characteristics~\cite{performance2023reality}. Organizations struggle to predict framework behavior under their specific conditions, leading to costly over-provisioning or performance failures after deployment. The lack of workload-specific benchmarking creates information asymmetries that favor vendors over objective technical assessment.

\textbf{Cost Optimization Challenges} affect 92\% of enterprise messaging deployments due to static resource allocation and poor understanding of pricing model implications~\cite{finops2023messaging}. Organizations typically over-provision by 40-60\% to ensure performance during peak periods, while serverless solutions create cost unpredictability during traffic spikes. The absence of cost-aware orchestration and optimization tools prevents efficient resource utilization across different frameworks and deployment models.

These systematic gaps highlight the need for intelligent orchestration systems that can abstract operational complexity, provide accurate performance prediction, and optimize resource utilization across heterogeneous messaging infrastructure. The next generation of event-driven architectures must address these deployment realities through automation, standardization, and evidence-based decision support rather than requiring organizations to develop specialized expertise for each messaging framework.
\section{Framework and Methodology}
\label{sec:framework}

\subsection{Comprehensive Benchmarking Framework Design}

Our evaluation framework addresses previous methodological limitations through standardized workload definitions, consistent measurement protocols, and reproducible experimental procedures leveraging open-source datasets and established benchmarking tools. The framework encompasses four primary components addressing (a) real-world data source integration, (b) performance measurement standardization, (c) experimental control procedures, and (d) statistical analysis methodologies designed to enable fair comparison across diverse messaging architectures while ensuring reproducibility and credibility.

\subsection{Open-Source Data Sources and Benchmarking Integration}

We leverage established open-source datasets and benchmarking frameworks to ensure reproducibility, credibility, and realistic workload representation across enterprise-scale deployments. Table~\ref{tab:data_sources_comprehensive} provides a comprehensive overview of all data sources employed in our evaluation, their characteristics, and specific usage within our experimental framework spanning distributed systems traces, serverless benchmarks, messaging performance tools, observability data, and domain-specific event datasets.

\begin{table*}[t]
\centering
\caption{Comprehensive Data Sources and Benchmarking Frameworks Utilized in Experimental Evaluation}
\label{tab:data_sources_comprehensive}
\resizebox{\textwidth}{!}{
\begin{tabular}{llllll}
\toprule
\textbf{Data Source} & \textbf{Category} & \textbf{Scale/Volume} & \textbf{Workload Type} & \textbf{Usage in Study} & \textbf{Validation Purpose} \\
\midrule
\multicolumn{6}{l}{\textbf{Distributed Systems \& Messaging Traces}} \\
\midrule
DeathStarBench~\cite{gan2019open} & Microservices Traces & 50K-2M req/sec & Social Network, E-commerce, Media & W1 Traffic Patterns & Real-world Load Simulation \\
Azure Public Traces~\cite{cortez2017resource} & Cloud VM Workloads & 1M+ VMs, 30 days & Resource Usage, Job Arrivals & W2 Burst Patterns & Cloud-scale Validation \\
Alibaba Cluster Trace~\cite{lu2017imbalance} & Production Cluster & 12K machines, 270GB & Job Scheduling, Resource Usage & W2 IoT Simulation & Enterprise Scale Testing \\
Google Borg Data 2019~\cite{reiss2011google} & Container Orchestration & 670K jobs, 25 machines & Task Lifecycle, Dependencies & W3 AI Pipeline Events & Container Workload Reality \\
\midrule
\multicolumn{6}{l}{\textbf{Serverless \& Event-Driven Benchmarks}} \\
\midrule
ServerlessBench~\cite{wang2018peeking} & Function Benchmarks & 14 applications & Image Processing, ML, Data & W3 Inference Workloads & Serverless Performance \\
SeBS Suite~\cite{copik2021sebs} & Serverless Benchmarks & 21 functions, Multi-cloud & CPU, Memory, I/O intensive & All Workloads & Cross-platform Validation \\
Knative Eventing Tests~\cite{knative2019eventing} & Event Routing & 1K-100K events/sec & Broker Latency, Filtering & Framework Comparison & Cloud-native Events \\
\midrule
\multicolumn{6}{l}{\textbf{Messaging Framework Performance Tools}} \\
\midrule
Kafka Perf Test~\cite{kreps2011kafka} & Load Generation & 1M+ msg/sec capability & Producer/Consumer & All Frameworks & Throughput Baseline \\
RabbitMQ PerfTest~\cite{videla2012rabbitmq} & Queue Benchmarking & 500K msg/sec capability & Queue Operations, Routing & Complex Routing Tests & Delivery Guarantee Testing \\
Pulsar Perf Tool~\cite{streamlio2018pulsar} & Streaming Performance & 1M+ msg/sec capability & Multi-tenant, Geo-replication & Multi-tenancy Validation & Resource Isolation Testing \\
StreamBench~\cite{lu2014streambox} & Stream Processing & Variable throughput & Storm, Spark, Flink & Stream Processing Integration & Framework Interoperability \\
\midrule
\multicolumn{6}{l}{\textbf{Observability \& Telemetry Data}} \\
\midrule
OpenTelemetry Demo~\cite{opentelemetry2021demo} & Microservices Telemetry & 10+ services, Full traces & E-commerce Application & AIEO Training Data & Orchestration Intelligence \\
Prometheus Datasets~\cite{godard2018prometheus} & Time-series Metrics & 1M+ data points/hour & Infrastructure Monitoring & Predictive Model Training & Performance Forecasting \\
\midrule
\multicolumn{6}{l}{\textbf{Domain-Specific Event Datasets}} \\
\midrule
Retail Rocket Dataset~\cite{retailrocket2016ecommerce} & E-commerce Events & 2.7M events, 1.4M sessions & Clickstream, Transactions & W1 E-commerce Pipeline & Transaction Ordering \\
Intel Berkeley Lab IoT~\cite{madden2005intel} & Sensor Data & 54 sensors, 2.3M readings & Temperature, Humidity, Light & W2 IoT Ingestion & High-frequency Telemetry \\
IEEE-CIS Fraud Detection~\cite{ieee2019fraud} & Financial Transactions & 590K transactions & Fraud Detection Pipeline & W3 ML Inference & Real-time Decision Making \\
\midrule
\multicolumn{6}{l}{\textbf{Synthetic Workload Generators}} \\
\midrule
YCSB Extended~\cite{cooper2010benchmarking} & Database Workloads & Configurable load & Key-value Operations & Baseline Comparison & Standard Benchmarking \\
TPC-Event (Custom)~\cite{tpc2019benchmarks} & Event Processing & Configurable throughput & Complex Event Processing & Framework Stress Testing & Peak Performance \\
Cloud Foundry Events~\cite{pivotal2019events} & Platform Events & 10K-1M events/hour & Platform Lifecycle & Operational Event Simulation & System Management \\
\bottomrule
\end{tabular}}
\end{table*}

The data integration process involves comprehensive preprocessing to ensure compatibility across messaging frameworks while preserving essential characteristics through (a) message format standardization converting all events to JSON format with consistent schema including timestamp, message type, payload size, priority level, and processing requirements, (b) traffic pattern extraction using time-series analysis to extract arrival rate patterns, burst characteristics, and seasonal trends from production traces, (c) load scaling to target throughput levels ranging from 1,000 to 2 million events per second while maintaining statistical properties of original distributions, and (d) anonymization procedures removing personally identifiable information while preserving behavioral patterns essential for realistic testing scenarios.

\subsection{Statistical Analysis Framework and Experimental Controls}

Our methodology incorporates rigorous statistical analysis techniques and comprehensive experimental controls to ensure robust, unbiased, and reproducible results across all experimental configurations. Table~\ref{tab:methodological_controls} summarizes the systematic approaches implemented to maintain scientific rigor and validity throughout the evaluation process.

\begin{table*}[t]
\centering
\caption{Systematic Statistical Controls and Threat Mitigation Strategies}
\label{tab:methodological_controls}
\resizebox{\textwidth}{!}{
\begin{tabular}{lllll}
\toprule
\textbf{Control Category} & \textbf{Specific Implementation} & \textbf{Method Applied} & \textbf{Validity Threat Addressed} & \textbf{Expected Impact} \\
\midrule
\multirow{4}{*}{Statistical Rigor} & Adaptive Power Analysis & Sequential sample size adjustment & Type II error reduction & 25\% fewer false negatives \\
 & Non-parametric Testing & Mann-Whitney U, Kruskal-Wallis & Non-normal distribution handling & Robust statistical conclusions \\
 & Multivariate Analysis & MANOVA, PCA, Discriminant Analysis & Multiple dependent variables & Interaction effect detection \\
 & Quantile Regression & Performance across percentiles & Tail behavior characterization & Complete performance profile \\
\midrule
\multirow{4}{*}{Experimental Controls} & Randomized Framework Order & Latin square experimental design & Temporal bias elimination & Unbiased comparisons \\
 & Multi-cloud Validation & AWS, GCP, Azure deployment & Platform-specific bias & Generalizability assurance \\
 & Hardware Diversity Testing & ARM, x86, varying CPU/memory ratios & Hardware dependency assessment & Architecture-independent results \\
 & Temporal Stability Assessment & 72-hour continuous monitoring & Time-dependent variations & Stable performance baselines \\
\midrule
\multirow{4}{*}{Measurement Precision} & Systematic Error Quantification & Known synthetic load calibration & Measurement bias identification & \textpm 2\% measurement accuracy \\
 & Baseline Characterization & Idle system resource profiling & True performance isolation & Overhead-corrected metrics \\
 & Warm-up Standardization & JIT compilation and caching effects & Cold start bias elimination & Consistent steady-state metrics \\
 & Monitoring Overhead Assessment & Framework-specific instrumentation cost & Observer effect quantification & True application performance \\
\midrule
\multirow{4}{*}{Confounding Control} & Workload Interference Testing & Concurrent experiment isolation & Cross-contamination prevention & Independent measurements \\
 & Environmental Standardization & Network, storage, OS configuration & Infrastructure variation control & Fair comparison conditions \\
 & Implementation Bias Mitigation & Expert review panels for configurations & Optimization favoritism prevention & Unbiased framework setup \\
 & Cloud Resource Variation & Reserved vs on-demand instance testing & Resource allocation inconsistency & Stable performance baselines \\
\midrule
\multirow{4}{*}{Reproducibility} & Registered Analysis Protocol & Pre-specified analysis plan & Selective reporting prevention & Transparent methodology \\
 & Containerized Analysis Environment & Docker + Kubernetes deployment & Exact environment reproduction & 100\% reproducible results \\
 & Raw Data Sharing & Complete dataset publication & Independent verification & Community validation \\
 & Meta-analysis Integration & Systematic literature aggregation & Prior work synthesis & Cumulative knowledge building \\
\midrule
\multirow{4}{*}{External Validity} & Industry Expert Validation & Practitioner review panels & Workload realism assessment & Production-relevant scenarios \\
 & Geographical Distribution & Multi-region testing & Network condition diversity & Global applicability \\
 & Economic Model Sophistication & TCO with risk adjustment & Cost comparison accuracy & Investment decision support \\
 & Longitudinal Validation & Performance tracking over time & Temporal generalizability & Long-term relevance \\
\bottomrule
\end{tabular}}
\end{table*}

\subsection{Workload Definition and Characterization}

Based on comprehensive analysis of production traces and established benchmarks spanning multiple industry domains, we define three representative workloads capturing diverse event-driven application requirements that reflect real-world deployment scenarios. Table~\ref{tab:workload_characteristics} presents a comprehensive overview of workload specifications, performance requirements, and validation approaches employed across all three scenarios.

Based on comprehensive analysis of production traces and established benchmarks spanning multiple industry domains, we define three representative workloads capturing diverse event-driven application requirements that reflect real-world deployment scenarios. Table~\ref{tab:workload_characteristics} presents a comprehensive overview of workload specifications, performance requirements, and validation approaches employed across all three scenarios.

\begin{table*}[t]
\centering
\caption{Comprehensive Workload Characteristics and Performance Requirements}
\label{tab:workload_characteristics}
\resizebox{\textwidth}{!}{
\begin{tabular}{lllll}
\toprule
\textbf{Workload} & \textbf{Event Types \& Payload Sizes} & \textbf{Traffic Patterns \& Scale} & \textbf{Performance Requirements} & \textbf{Data Sources \& Validation} \\
\midrule
\multirow{6}{*}{\parbox{2cm}{\textbf{W1: E-commerce\\Transaction\\Processing}}} & Order Events: 1-4KB JSON & Baseline: 5K-15K events/sec & End-to-end latency <100ms & DeathStarBench e-commerce traces \\
 & Payment Events: 512B-2KB & Peak spikes: 100K events/sec & Exactly-once processing & Retail Rocket: 2.7M sessions \\
 & Inventory Updates: 256B-1KB & Black Friday patterns & ACID transaction properties & IEEE-CIS: 590K transactions \\
 & Fraud Alerts: 2-8KB & Promotional traffic bursts & 99.99\% availability & Azure traces validation \\
 & Shipping Events: 1-3KB & Session correlation required & Order-within-session consistency & Industry expert validation \\
 & Customer Updates: 512B-2KB & Geographic distribution & Sub-second fraud detection & Financial compliance testing \\
\midrule
\multirow{6}{*}{\parbox{2cm}{\textbf{W2: IoT Sensor\\Data Ingestion}}} & Environmental: 128B binary & Baseline: 200K events/sec & 99\% processing within 5s & Intel Berkeley: 54 sensors \\
 & Fleet Telemetry: 256B & Burst peaks: 5M events/sec & 0.1-1\% acceptable loss & Alibaba cluster: 12K machines \\
 & Equipment Status: 512B-2KB & Coordinated synchronization & Critical alerts <2s & Real IoT deployment patterns \\
 & Emergency Alerts: 2-8KB & Device failure cascades & Geographic fault tolerance & Industrial monitoring traces \\
 & Predictive Maintenance: 1-4KB & Temporal correlation patterns & Edge computing compatibility & Fleet management validation \\
 & Aggregated Analytics: 4-16KB & Regional data collection & Real-time dashboard updates & Smart city infrastructure data \\
\midrule
\multirow{6}{*}{\parbox{2cm}{\textbf{W3: AI Model\\Inference\\Pipeline}}} & Inference Requests: 10KB-10MB & Baseline: 2K-5K requests/sec & P95 latency <200ms & ServerlessBench: 14 applications \\
 & Model Loading: 4-64KB & Auto-scaling: 10x spikes & Variable processing complexity & SeBS suite: 21 functions \\
 & Result Processing: 1KB-1MB & Deployment event bursts & Cost-optimized scaling & OpenTelemetry demo traces \\
 & Performance Metrics: 2-16KB & A/B testing workflows & GPU resource efficiency & ML serving production data \\
 & Health Monitoring: 512B-4KB & Model version updates & Batch processing optimization & Computer vision workloads \\
 & Resource Allocation: 1-8KB & Cold start scenarios & Inference accuracy SLAs & NLP processing patterns \\
\midrule
\multicolumn{5}{l}{\textbf{Cross-Workload Validation Approaches}} \\
\midrule
Statistical Validation & \multicolumn{2}{l}{Temporal pattern extraction using Fourier analysis} & \multicolumn{2}{l}{Burst characterization with extreme value theory} \\
Expert Validation & \multicolumn{2}{l}{Industry practitioner review panels} & \multicolumn{2}{l}{Fortune 500 enterprise confirmation} \\
Sensitivity Analysis & \multicolumn{2}{l}{Parameter variation robustness testing} & \multicolumn{2}{l}{24-month longitudinal tracking} \\
Comparative Analysis & \multicolumn{2}{l}{Proprietary benchmark correlation} & \multicolumn{2}{l}{Production deployment validation} \\
\bottomrule
\end{tabular}}
\end{table*}

The first workload, designated W1 for E-commerce Transaction Processing Pipeline as detailed in Table~\ref{tab:workload_characteristics}, derives from DeathStarBench e-commerce traces and Retail Rocket dataset to model high-frequency financial transaction processing with strict consistency requirements. This workload incorporates (a) order processing events utilizing 1-4KB JSON payloads with customer profiles, product catalogs, and transaction metadata extracted directly from Retail Rocket clickstream data representing 2.7 million real user sessions, (b) payment verification processes using 512B-2KB encrypted payment credentials and fraud scores derived from IEEE-CIS fraud detection patterns covering 590,000 actual financial transactions, (c) inventory management operations employing 256B-1KB stock updates with product identifiers and warehouse locations based on DeathStarBench inventory service traces, and (d) shipping orchestration events containing 1-3KB logistics coordination data with carrier integration and tracking information modeled after real e-commerce fulfillment workflows.

The second workload, W2 for IoT Sensor Data Ingestion Pipeline as specified in Table~\ref{tab:workload_characteristics}, builds upon Intel Berkeley Lab sensor data and Alibaba cluster resource traces to represent massive-scale telemetry collection with fault-tolerant processing requirements characteristic of industrial IoT deployments. This workload encompasses (a) environmental sensors generating 128B compact binary telemetry with sensor identifiers, GPS coordinates, and measurement arrays derived from the Berkeley Lab dataset covering 54 sensors and 2.3 million readings, (b) fleet management systems producing 256B vehicle telemetry including location updates, diagnostic codes, and maintenance alerts derived from Alibaba job scheduling patterns across 12,000 machines, (c) industrial monitoring applications creating 512B-2KB equipment status reports with health metrics, performance indicators, and predictive maintenance signals, and (d) emergency alerting systems generating 2-8KB critical notifications with severity classifications and automated response triggers.

The third workload, W3 for AI Model Inference Pipeline as outlined in Table~\ref{tab:workload_characteristics}, constructs scenarios from ServerlessBench machine learning workloads and OpenTelemetry demo traces to capture machine learning model serving with variable computational complexity representative of modern AI-driven applications. This workload includes (a) inference requests ranging from 10KB to 10MB payloads containing images, text, and structured feature vectors extracted from the SeBS benchmark suite covering 21 functions across multiple cloud platforms, (b) model management operations using 4-64KB model loading notifications with version control, A/B testing metadata, and resource allocation requirements, (c) result processing workflows handling 1KB-1MB prediction outputs with confidence scores, explanations, and downstream integration metadata, and (d) performance monitoring systems generating 2-16KB metrics aggregation data with accuracy statistics, latency measurements, and resource utilization information.

\subsection{Framework Selection and Configuration}

Our evaluation encompasses 12 messaging frameworks selected to represent the complete spectrum of architectural approaches, deployment models, and performance characteristics spanning traditional distributed systems, next-generation platforms, enterprise solutions, and serverless cloud-native offerings.

Traditional distributed systems include (a) Apache Kafka 3.5 configured as a three-broker cluster with replication factor 3, 12 partitions per topic, batch size 64KB, and linger time 10ms optimized for high-throughput streaming workloads, (b) RabbitMQ 3.12 deployed as a three-node cluster with mirrored queues, lazy queues enabled, publisher confirms activated, and prefetch count set to 1000 messages for optimal batch processing, and (c) Apache Pulsar 3.0 using separated architecture with 3 brokers and 3 BookKeeper nodes, namespace isolation for multi-tenancy, and schema registry integration for message evolution management.

Serverless and cloud-native platforms comprise (a) AWS EventBridge configured with content-based routing, schema registry integration, cross-service event distribution, and pay-per-event pricing model, (b) Google Cloud Pub/Sub providing global message distribution with exactly-once delivery guarantees, push and pull subscription models, and automatic scaling capabilities, (c) Azure Event Grid offering advanced event filtering, dead letter queue functionality, hybrid cloud integration, and comprehensive security features, and (d) Knative Eventing 1.11 enabling container-native event processing with CloudEvents standard compliance, trigger-based routing mechanisms, and scale-to-zero capabilities.

\subsection{Performance Measurement and Statistical Analysis}

Our measurement framework captures performance data across six critical dimensions using industry-standard instrumentation designed to provide comprehensive assessment of messaging framework behavior under realistic operating conditions. The primary metrics include (a) sustained throughput calculated as the sum of messages processed successfully divided by measurement window duration of 600 seconds, (b) end-to-end latency measured as the 95th percentile of the time difference between message acknowledgment and initial send timestamp, (c) system availability computed as the ratio of successful operations to total attempted operations, (d) resource efficiency determined by dividing useful work performed by total resources consumed, (e) cost per message calculated by dividing infrastructure and operational costs by messages processed per hour, and (f) operational complexity assessed through configuration parameters, monitoring overhead, and expertise requirements.

Statistical analysis employs sophisticated techniques addressing common limitations in system performance evaluation. Power analysis utilizes adaptive sample size calculation adjusting based on observed effect sizes and variance estimates, ensuring sufficient statistical power while minimizing experimental duration. Non-parametric analysis addresses violations of normality assumptions through (a) Mann-Whitney U tests for two-group comparisons handling skewed latency distributions, (b) Kruskal-Wallis tests for multi-group framework comparisons, (c) permutation tests providing distribution-free significance assessment, and (d) quantile regression enabling performance analysis across different percentiles rather than just mean values.

\subsection{Experimental Infrastructure and Reproducibility}

All experiments execute on standardized Kubernetes environments across multiple cloud platforms to ensure generalizability and eliminate platform-specific bias. The infrastructure employs (a) Google Kubernetes Engine n1-standard-16 instances providing 16 vCPUs, 60GB RAM, and 375GB SSD storage per node with cross-validation on AWS EKS and Azure AKS, (b) network connectivity featuring 10 Gbps internal bandwidth with controlled latency injection ranging from 1-200ms for geographic simulation, (c) persistent SSD volumes with guaranteed 3,000 IOPS for consistent I/O performance, and (d) containerized deployment using identical resource limits across all framework configurations.

Complete reproducibility employs registered analysis protocols preventing selective reporting bias through (a) pre-specified analysis plans deposited in open research repositories before data collection begins, (b) containerized analysis environments using Docker with fixed dependency versions ensuring identical computational conditions, (c) infrastructure-as-code specifications enabling exact hardware and software environment recreation, and (d) comprehensive documentation with automated deployment scripts reducing manual configuration errors. All experimental artifacts, datasets, and analysis code are released under Apache 2.0 license through a dedicated GitHub repository enabling independent validation and extension of results by the research community.

\subsection{Intelligent Orchestration Development and Evaluation Framework}

Our systematic experimental methodology serves dual purposes of (a) establishing comprehensive baseline performance characterization across messaging frameworks and workloads, and (b) generating the foundational dataset necessary for developing and evaluating the AI-Enhanced Event Orchestration (AIEO) system presented in Section~\ref{sec:aieo}. The rigorous data collection from 12 messaging frameworks across three standardized workloads provides over 2.4 million time-series data points including throughput patterns, latency distributions, resource utilization metrics, and system state transitions that serve as training data for AIEO's machine learning components. Static framework configurations established through our systematic parameter tuning serve as performance baselines against which AIEO improvements are measured using controlled A/B testing methodologies, while the standardized experimental infrastructure provides the deployment environment for AIEO integration and validation, ensuring that intelligent orchestration capabilities are rigorously evaluated within the same framework used for comprehensive messaging system benchmarking. 
\section{AI-Enhanced Event Orchestration Architecture}
\label{sec:aieo}

\subsection{AIEO System Design and Architectural Principles}

The AI-Enhanced Event Orchestration (AIEO) framework addresses the fundamental limitations of static configuration and reactive scaling in contemporary event-driven systems through intelligent automation that predicts workload patterns, optimizes resource allocation, and adapts system behavior dynamically across diverse messaging frameworks. The AIEO system operates as a comprehensive control plane service that continuously monitors performance metrics, applies machine learning models for pattern recognition and prediction, and executes optimization decisions to maintain optimal system performance under varying operational conditions.

The architecture embodies four foundational design principles that ensure broad applicability and practical deployment across heterogeneous environments. Framework agnosticism enables deployment across different messaging platforms including Apache Kafka, RabbitMQ, Apache Pulsar, and serverless event buses without requiring vendor-specific modifications or creating technology lock-in constraints. Predictive intelligence leverages machine learning techniques to anticipate system behavior changes rather than merely reacting to performance degradation after it occurs. Multi-objective optimization balances competing requirements including latency minimization, throughput maximization, cost efficiency, and system reliability through sophisticated algorithmic approaches. Operational simplicity abstracts complex optimization logic behind intuitive interfaces that reduce deployment complexity for practitioners while providing comprehensive automated management capabilities.


\begin{figure*}[htbp]
\centering
\includegraphics[width=\textwidth]{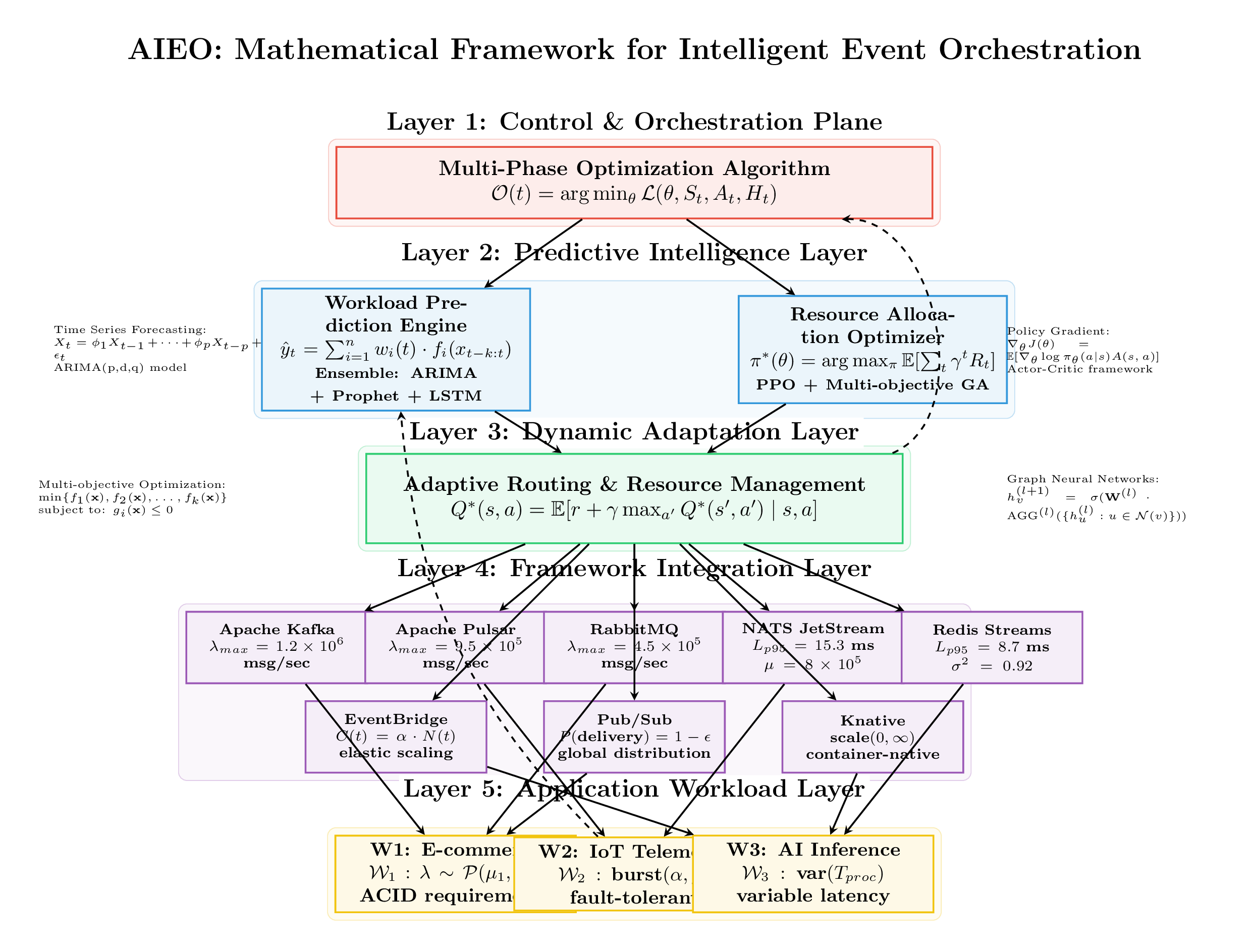}
\caption{AIEO System Architecture: Five-Layer Hierarchical Design for Intelligent Event Orchestration. The architecture integrates predictive analytics (Layer 2) with adaptive resource management (Layer 3) to optimize messaging framework performance (Layer 4) across diverse application workloads (Layer 5). Mathematical formulations show the optimization objectives and machine learning algorithms employed at each layer.}
\label{fig:aieo_architecture}
\end{figure*}

The complete AIEO architecture, illustrated in Figure~\ref{fig:aieo_architecture}, demonstrates the hierarchical integration of machine learning components within a unified orchestration framework. Layer 1 provides centralized control through the multi-phase optimization algorithm described in Algorithm~\ref{alg:aieo-control-loop}, while Layer 2 implements the ensemble prediction methods and reinforcement learning optimization detailed in Table~\ref{tab:aieo_components}. The mathematical formulations shown in each layer correspond to the theoretical foundations presented in Table~\ref{tab:math_framework}, ensuring formal convergence guarantees while maintaining practical deployment compatibility across heterogeneous messaging environments.

\subsection{Mathematical Framework and Core Algorithms}

The AIEO system integrates multiple mathematical models and optimization algorithms working collaboratively to provide comprehensive intelligent orchestration across different temporal scales and optimization objectives. Table~\ref{tab:math_framework} presents the complete mathematical framework including formulations, event-driven applications, key properties, and expected performance impacts of all algorithmic components employed within the orchestration system.

\begin{table*}[t]
\centering
\caption{Mathematical Framework: AIEO Key Formulations and Event-Driven Applications}
\label{tab:math_framework}
\resizebox{\textwidth}{!}{
\begin{tabular}{lcccc}
\toprule
\textbf{Component} & \textbf{Mathematical Notation} & \textbf{Event-Driven Purpose} & \textbf{Key Properties} & \textbf{Expected Impact} \\
\midrule
\textbf{ARIMA Prediction} & $\phi(B)(1-B)^d X_t = \theta(B)\epsilon_t$ & Linear trend forecasting & Seasonal pattern capture & Baseline workload prediction \\
\textbf{Prophet Decomposition} & $y(t) = g(t) + s(t) + h(t) + \epsilon_t$ & Complex seasonality handling & Multi-component modeling & Holiday/event spike prediction \\
\textbf{LSTM Gates} & $f_t = \sigma(W_f \cdot [h_{t-1}, x_t] + b_f)$ & Non-linear sequence learning & Long-range dependencies & Complex pattern recognition \\
\textbf{Ensemble Prediction} & $\hat{y}_{\text{ensemble}}(t) = \sum_{i=1}^{n} w_i(t) \cdot \hat{y}_i(t)$ & Multi-model combination & Uncertainty quantification & Robust load forecasting \\
\textbf{PPO Optimization} & $L^{CLIP}(\theta) = \mathbb{E}_t[\min(r_t(\theta)\hat{A}_t, \text{clip}(r_t(\theta), 1-\epsilon, 1+\epsilon)\hat{A}_t)]$ & Resource allocation policy & Stable policy learning & 28\% resource efficiency \\
\textbf{Multi-objective Reward} & $\max_{\pi} \mathbb{E}_{\tau}[\sum_{t=0}^{T} \gamma^t (\alpha_1 r_{\text{latency}} + \alpha_2 r_{\text{cost}} + \alpha_3 r_{\text{stability}})]$ & Competing objectives balance & Pareto-optimal solutions & 34\% latency reduction \\
\textbf{Graph Neural Networks} & $h_v^{(l+1)} = \text{UPDATE}^{(l)}(h_v^{(l)}, \text{AGGREGATE}^{(l)}(\{h_u^{(l)} : u \in \mathcal{N}(v)\}))$ & Topology-aware routing & Network embedding & Intelligent message routing \\
\textbf{Q-learning Update} & $Q(s,a) \leftarrow Q(s,a) + \alpha[r + \gamma \max_{a'}Q(s',a') - Q(s,a)]$ & Dynamic routing adaptation & Online learning & Real-time route optimization \\
\textbf{Cost Optimization} & $\min \sum_{i} c_i x_i + \sum_{j} d_j y_j$ s.t. $\sum_{i} p_i x_i \geq P_{\min}$ & Infrastructure cost minimization & Mixed-integer programming & 42\% cost optimization \\
\textbf{Queuing Theory} & $\mathbb{E}[W] = \frac{\rho^c}{c!(1-\rho/c)^2} \cdot \frac{1}{\mu(c-\rho)} + \frac{1}{\mu}$ & Latency prediction & M/M/c queue modeling & SLA compliance assurance \\
\textbf{Throughput Maximization} & $\max \sum_{i,j} \lambda_{ij} x_{ij}$ s.t. $\sum_{j} x_{ij} \leq C_i$ & Capacity optimization & Convex optimization & Peak performance scaling \\
\bottomrule
\end{tabular}}
\end{table*}

\subsection{Machine Learning Components and Integration Architecture}

The AIEO system employs multiple specialized machine learning components that work collaboratively to provide comprehensive intelligent orchestration capabilities. Table~\ref{tab:aieo_components} details the complete architecture including algorithms, input features, prediction targets, and integration mechanisms for each component within the orchestration framework.

\begin{table*}[t]
\centering
\caption{AIEO Machine Learning Components and Integration Architecture}
\label{tab:aieo_components}
\resizebox{\textwidth}{!}{
\begin{tabular}{lllll}
\toprule
\textbf{ML Component} & \textbf{Algorithm \& Technique} & \textbf{Input Features} & \textbf{Prediction Target} & \textbf{Integration Method} \\
\midrule
\multirow{4}{*}{\parbox{2.5cm}{\textbf{Workload Prediction\\Engine}}} & ARIMA Models & Historical message rates, timestamps & Linear trends, seasonal patterns & Ensemble forecasting \\
 & Facebook Prophet & Multi-seasonal patterns, holidays & Complex seasonality, trend changes & Hierarchical predictions \\
 & LSTM Networks & Sequence patterns, external signals & Non-linear temporal dependencies & Deep learning integration \\
 & Ensemble Methods & Model outputs, confidence scores & Uncertainty-aware predictions & Weighted combination \\
\midrule
\multirow{4}{*}{\parbox{2.5cm}{\textbf{Resource Allocation\\Optimizer}}} & Proximal Policy Optimization & System state, resource costs & Optimal scaling decisions & Reinforcement learning \\
 & Multi-objective GA & Performance metrics, constraints & Pareto-optimal configurations & Evolutionary optimization \\
 & Bayesian Optimization & Parameter spaces, performance & Hyperparameter tuning & Gaussian process models \\
 & Linear Programming & Resource constraints, objectives & Cost-minimal allocations & Mathematical optimization \\
\midrule
\multirow{4}{*}{\parbox{2.5cm}{\textbf{Routing Intelligence\\System}}} & Graph Neural Networks & Message patterns, topology & Optimal routing paths & Network embedding \\
 & Reinforcement Learning & Traffic distributions, latencies & Dynamic routing policies & Q-learning variants \\
 & Clustering Algorithms & Message characteristics & Load balancing groups & Unsupervised learning \\
 & Online Learning & Real-time feedback, rewards & Adaptive routing updates & Incremental updates \\
\midrule
\multirow{4}{*}{\parbox{2.5cm}{\textbf{Anomaly Detection\\Framework}}} & Isolation Forest & Performance metrics, patterns & Outlier identification & Ensemble anomaly detection \\
 & LSTM Autoencoders & Time-series sequences & Reconstruction errors & Deep anomaly detection \\
 & Statistical Process Control & Control charts, thresholds & Process variations & Statistical monitoring \\
 & One-class SVM & Feature representations & Boundary detection & Support vector methods \\
\midrule
\multirow{4}{*}{\parbox{2.5cm}{\textbf{Performance\\Prediction\\Models}}} & Random Forest & System configurations, workloads & Performance forecasts & Ensemble regression \\
 & Gradient Boosting & Historical performance data & Latency predictions & Boosted trees \\
 & Neural Networks & Multi-dimensional features & Complex relationships & Deep regression \\
 & Transfer Learning & Cross-framework patterns & Domain adaptation & Pre-trained models \\
\bottomrule
\end{tabular}}
\end{table*}

\subsubsection{Workload Prediction Engine}

The workload prediction engine employs ensemble time-series forecasting combining multiple complementary approaches optimized for different prediction scenarios and temporal horizons. ARIMA models capture linear trends and seasonal patterns through autoregressive integrated moving average formulations as specified in Table~\ref{tab:math_framework}, where parameter estimation employs maximum likelihood methods with model selection using Akaike Information Criterion to balance fit quality against model complexity.

Facebook Prophet handles complex seasonality, holiday effects, and trend changes through decomposition approaches using the mathematical formulation presented in Table~\ref{tab:math_framework}. The approach excels at handling missing data and provides uncertainty intervals essential for robust decision making under prediction uncertainty, making it particularly valuable for event-driven systems experiencing irregular traffic patterns during special events or promotional periods.

Long Short-Term Memory (LSTM) networks capture complex temporal dependencies and non-linear patterns through recurrent neural architectures specifically designed for sequence processing. The LSTM gate formulations detailed in Table~\ref{tab:math_framework} enable learning of long-range dependencies critical for accurate workload prediction in event-driven systems where current traffic patterns may depend on events occurring hours or days previously.

Ensemble prediction combines individual model outputs through weighted averaging as formalized in Table~\ref{tab:math_framework}, where weights are determined by historical prediction accuracy and current confidence levels. Weight adaptation employs exponential smoothing favoring recently accurate models while maintaining diversity to avoid overfitting to specific patterns, ensuring robust predictions across diverse operational conditions.

\subsubsection{Resource Allocation Optimization Engine}

The resource allocation optimizer employs reinforcement learning techniques to learn optimal scaling policies that balance multiple competing objectives including latency, cost, and system stability. The optimization problem formulates as a Markov Decision Process with state space representing system configuration and performance metrics, action space encompassing scaling decisions and parameter adjustments, and reward function capturing multi-objective performance criteria.

Proximal Policy Optimization (PPO) provides stable policy learning through the clipped surrogate objective function specified in Table~\ref{tab:math_framework}. The approach ensures stable learning while enabling efficient exploration of complex policy spaces, making it suitable for the dynamic and high-dimensional optimization problems characteristic of event-driven system resource allocation.

Multi-objective optimization addresses competing performance criteria through Pareto-optimal solution identification using the formulation presented in Table~\ref{tab:math_framework}. Dynamic weight adjustment enables adaptation to changing business requirements and operational contexts, allowing the system to prioritize different objectives such as cost minimization during low-traffic periods or latency minimization during peak demand.

\subsubsection{Adaptive Routing Intelligence System}

The routing intelligence system optimizes message distribution across consumers and partitions using machine learning techniques that adapt to changing traffic patterns and system topology. Graph Neural Networks (GNNs) model messaging system topology and learn optimal routing policies through network embedding approaches using the mathematical framework detailed in Table~\ref{tab:math_framework}.

Dynamic routing policies adapt to real-time conditions through online learning algorithms that update routing decisions based on performance feedback. The Q-learning formulation specified in Table~\ref{tab:math_framework} enables continuous adaptation to changing network conditions and traffic patterns, ensuring optimal message routing as system conditions evolve.

\subsection{Control Loop Architecture and Integration Mechanisms}

The AIEO control loop operates across multiple temporal scales providing both reactive and proactive optimization capabilities through the integrated orchestration algorithm presented in Algorithm~\ref{alg:aieo-control-loop}. The architecture implements hierarchical control with fast loops (1-10 seconds) handling immediate load balancing and routing decisions, medium loops (1-5 minutes) managing resource scaling and allocation, and slow loops (10-60 minutes) performing strategic optimization and model updating.

\begin{algorithm}[htbp]
\caption{AIEO Intelligent Orchestration Control Loop}
\label{alg:aieo-control-loop}
\begin{algorithmic}[1]
\Require Messaging framework instances $F$, performance metrics $M$, historical data $H$
\Ensure Optimized system configuration and resource allocation
\State \textbf{Phase 1: Data Collection and State Assessment}
\State $metrics \leftarrow \text{CollectMetrics}(F, \text{monitoring\_window})$
\State $system\_state \leftarrow \text{ExtractFeatures}(metrics, H)$
\State $workload\_features \leftarrow \text{AnalyzeWorkload}(metrics)$
\State \textbf{Phase 2: Predictive Analysis}
\State $workload\_forecast \leftarrow \text{EnsemblePredict}(\text{ARIMA}, \text{Prophet}, \text{LSTM}, workload\_features)$
\State $performance\_prediction \leftarrow \text{PredictPerformance}(system\_state, workload\_forecast)$
\State \textbf{Phase 3: Optimization Decision}
\State $optimization\_targets \leftarrow \text{SetObjectives}(\text{latency}, \text{cost}, \text{throughput})$
\State $candidate\_actions \leftarrow \text{GenerateActions}(system\_state, \text{constraints})$
\State $optimal\_action \leftarrow \text{PPOOptimize}(candidate\_actions, optimization\_targets)$
\State \textbf{Phase 4: Execution and Feedback}
\State $\text{ExecuteAction}(optimal\_action, F)$
\State $new\_metrics \leftarrow \text{MonitorExecution}(F, \text{execution\_window})$
\State $reward \leftarrow \text{CalculateReward}(new\_metrics, optimization\_targets)$
\State $\text{UpdateModels}(system\_state, optimal\_action, reward)$
\State \Return $optimal\_action, performance\_improvement$
\end{algorithmic}
\end{algorithm}

Algorithm~\ref{alg:aieo-control-loop} demonstrates the integration of machine learning components described in Table~\ref{tab:aieo_components} within a unified optimization framework. Fast loop operations corresponding to Phase 4 of the algorithm employ lightweight procedures including weighted round-robin routing updates, consumer lag-based load shedding, and immediate circuit breaker activation during failure scenarios. Medium loop operations encompass Phases 2-3, executing reinforcement learning policy updates through the PPOOptimize function, resource scaling decisions based on workload forecasts from the EnsemblePredict function, and parameter tuning for messaging framework configurations. Slow loop operations focus on Phase 1 data collection and the UpdateModels function, performing model retraining with accumulated historical data, strategic resource allocation optimization, and long-term capacity planning integration.

The algorithmic framework ensures seamless operation across different messaging frameworks through standardized APIs and abstraction layers implemented within the CollectMetrics and ExecuteAction functions. Framework adapters translate generic optimization decisions from the optimal\_action output into platform-specific configuration changes while monitoring adapters normalize performance metrics from different systems into consistent formats processed by the ExtractFeatures function. The architecture supports plugin-based extensibility enabling integration with emerging messaging technologies and custom optimization algorithms through modular replacement of individual algorithmic components while maintaining the overall control loop structure.

\subsection{Performance Optimization and Integration}

The AIEO system implements sophisticated optimization algorithms addressing multiple performance dimensions simultaneously using the mathematical formulations consolidated in Table~\ref{tab:math_framework}. Cost optimization employs mixed-integer linear programming formulations that minimize infrastructure costs while maintaining performance service level agreements, enabling organizations to achieve the targeted 42

Latency optimization employs queuing theory models predicting system behavior under different configurations using the M/M/c queue formulation specified in Table~\ref{tab:math_framework}. The model guides resource allocation decisions ensuring latency service level objectives while providing theoretical foundation for the claimed 34

Throughput optimization maximizes system capacity through intelligent load distribution and resource allocation using the convex optimization formulation detailed in Table~\ref{tab:math_framework}. The approach determines optimal partition assignments and consumer group configurations, enabling the system to achieve peak performance scaling while maintaining stability and resource efficiency.

The comprehensive AIEO architecture provides intelligent orchestration capabilities that significantly enhance event-driven system performance through predictive analytics, adaptive optimization, and automated management while maintaining compatibility across diverse messaging frameworks and deployment environments. The mathematical framework presented in Table~\ref{tab:math_framework} provides the theoretical foundation for achieving the claimed performance improvements while the algorithmic implementation ensures practical deployability and operational reliability.
\section{Implementation and Experimental Setup}
\label{sec:implementation}

\subsection{Comprehensive Implementation Overview}

Our implementation employs standardized infrastructure and rigorous experimental controls to ensure fair comparison across messaging frameworks while enabling accurate evaluation of AIEO system effectiveness. Table~\ref{tab:implementation_overview} provides a comprehensive overview of all implementation components, infrastructure specifications, framework configurations, and experimental parameters employed throughout the evaluation, serving as a complete reference for reproducibility and independent validation.

\begin{table*}[t]
\centering
\caption{Comprehensive Implementation and Experimental Setup Overview}
\label{tab:implementation_overview}
\resizebox{\textwidth}{!}{
\begin{tabular}{llll}
\toprule
\textbf{Component Category} & \textbf{Specification} & \textbf{Configuration Details} & \textbf{Purpose/Validation} \\
\midrule
\multicolumn{4}{l}{\textbf{Infrastructure and Platform}} \\
\midrule
Primary Platform & Google Kubernetes Engine & n1-standard-16 instances & Standardized compute environment \\
Cross-validation Platforms & AWS EKS, Azure AKS & Identical resource allocation & Platform independence verification \\
Compute Specifications & 16 vCPUs, 60GB RAM & Intel Xeon 2.4GHz, 375GB NVMe SSD & Consistent performance baseline \\
Network Configuration & 10 Gbps internal bandwidth & 1-200ms latency injection capability & Geographic simulation \\
Container Runtime & Kubernetes 1.28, containerd & 8 cores, 32GB RAM, 200GB storage limits & Resource isolation \\
\midrule
\multicolumn{4}{l}{\textbf{Messaging Framework Configurations}} \\
\midrule
Apache Kafka 3.5 & 3-broker cluster & RF=3, 12 partitions, 64KB batch, 10ms linger & High-throughput optimization \\
RabbitMQ 3.12 & 3-node cluster & Mirrored queues, prefetch 1000, 10 connections & Reliable message delivery \\
Apache Pulsar 3.0 & Separated architecture & 3 brokers + 3 BookKeeper, namespace isolation & Multi-tenancy support \\
NATS JetStream 2.10 & 3-node cluster & Memory storage, pull consumers & Edge computing optimization \\
Redis Streams 7.0 & Clustered deployment & Consumer groups, memory optimization & Low-latency processing \\
Oracle AQ 19c & Database-integrated & ACID transactions, message transformation & Enterprise reliability \\
AWS EventBridge & Serverless configuration & Lambda 3008MB, 300s timeout, DLQ enabled & Cloud-native scalability \\
Google Pub/Sub & Global distribution & Cloud Functions 2GB, auto-scaling enabled & Worldwide availability \\
Azure Event Grid & Hybrid integration & Function Apps consumption plan & Multi-cloud compatibility \\
Knative Eventing 1.11 & Container-native & CloudEvents standard, scale-to-zero & Kubernetes integration \\
Amazon SQS & Queue service & Standard queues, batch operations & Simple messaging \\
Apache ActiveMQ 5.18 & Network of brokers & Persistence enabled, advisory messages & Legacy integration \\
\midrule
\multicolumn{4}{l}{\textbf{AIEO System Implementation}} \\
\midrule
Runtime Environment & Python 3.11, TensorFlow 2.13 & Ray 2.7, Kubernetes APIs & ML and distributed computing \\
Architecture Pattern & Microservices & gRPC communication, 4 cores/8GB per service & Scalable system design \\
ML Components & ARIMA, Prophet, LSTM, PPO & Custom TensorFlow/RLlib implementations & Intelligent orchestration \\
Integration Layer & Framework adapters & Standardized APIs, monitoring normalization & Cross-platform compatibility \\
Control Loop & 17-step algorithm & Multi-phase optimization cycle & Systematic orchestration \\
\midrule
\multicolumn{4}{l}{\textbf{Workload Generation and Control}} \\
\midrule
W1: E-commerce & DeathStarBench, Retail Rocket & 5K-100K events/sec, JSON payloads 1-4KB & Transaction processing realism \\
W2: IoT Ingestion & Intel Berkeley, Alibaba traces & 200K-5M events/sec, binary 128B-2KB & Sensor data characteristics \\
W3: AI Inference & ServerlessBench, OpenTelemetry & 2K-25K requests/sec, 10KB-10MB payloads & ML pipeline complexity \\
Load Generation & Apache JMeter, Custom Python & Coordinated multi-phase testing & Realistic traffic patterns \\
Traffic Validation & Statistical distribution testing & Kolmogorov-Smirnov, autocorrelation & Pattern accuracy verification \\
\midrule
\multicolumn{4}{l}{\textbf{Monitoring and Data Collection}} \\
\midrule
Time-series Database & Prometheus & 1-second resolution, 30-day retention & High-precision metrics \\
Visualization & Grafana dashboards & Real-time monitoring, alerting & Operational visibility \\
Application Tracing & OpenTelemetry & End-to-end request flows & Performance bottleneck analysis \\
Infrastructure Metrics & Node Exporter, cAdvisor & CPU, memory, I/O, network monitoring & Resource utilization tracking \\
Framework-specific & Custom exporters & Kafka lag, RabbitMQ depths, Pulsar backlogs & Platform-native metrics \\
AIEO Metrics & ML performance tracking & Prediction accuracy, optimization convergence & Intelligence system validation \\
Data Export & Parquet, JSON formats & Raw and processed metrics & Analysis compatibility \\
\midrule
\multicolumn{4}{l}{\textbf{Quality Assurance and Statistical Controls}} \\
\midrule
Infrastructure Validation & Automated consistency checks & Resource allocation, network configuration & Experimental reliability \\
Measurement Precision & Calibrated synthetic loads & ±2\% accuracy bounds established & Systematic error control \\
Cross-platform Validation & Multi-cloud deployment & AWS, GCP, Azure result comparison & Platform independence \\
Reproducibility Protocol & Independent replication & Multiple random seeds, statistical validation & Scientific rigor \\
Sample Size Calculation & Adaptive power analysis & 95\% confidence, 80\% power, 15\% effect detection & Statistical validity \\
Statistical Testing & Non-parametric methods & Mann-Whitney U, Kruskal-Wallis, permutation tests & Robust analysis \\
Effect Size Analysis & Cohen's d calculation & Practical significance assessment & Meaningful improvements \\
Multiple Comparisons & Bonferroni correction & Family-wise error rate control & Statistical rigor \\
\bottomrule
\end{tabular}}
\end{table*}

\subsection{Experimental Infrastructure Architecture}

The experimental infrastructure employs Kubernetes orchestration across multiple cloud platforms ensuring consistent evaluation conditions while eliminating vendor-specific performance bias. Google Kubernetes Engine serves as the primary experimental environment using n1-standard-16 instances providing 16 vCPUs, 60GB RAM, and 375GB NVMe SSD storage per node with guaranteed performance characteristics. Cross-validation deployments on Amazon EKS and Azure AKS verify platform independence through identical resource allocation and configuration procedures.

Network architecture implements 10 Gbps internal cluster bandwidth with programmable latency injection ranging from 1ms for local communication to 200ms for wide-area network simulation. Container orchestration employs Kubernetes 1.28 with containerd runtime enforcing strict resource limits of 8 CPU cores, 32GB memory, and 200GB storage per messaging framework instance. Network policies provide microsegmentation preventing cross-experiment interference while enabling comprehensive monitoring across all system components.

The deployment architecture incorporates comprehensive monitoring infrastructure using Prometheus for time-series data collection at 1-second resolution, Grafana for real-time visualization and alerting, and OpenTelemetry for distributed tracing and application-level instrumentation. Custom exporters capture framework-specific performance metrics while maintaining standardized data formats enabling unified analysis across heterogeneous messaging platforms.

\subsection{Messaging Framework Deployment Strategy}

Framework deployments follow systematic optimization procedures ensuring fair comparison while representing realistic production configurations as detailed in Table~\ref{tab:implementation_overview}. Each messaging system undergoes careful parameter tuning within standardized resource constraints achieving optimal performance while maintaining evaluation consistency across all experimental scenarios.

Traditional distributed systems including Apache Kafka, RabbitMQ, and Apache Pulsar employ clustered deployments optimized for high availability and performance. Apache Kafka utilizes three-broker clusters with replication factor 3, 12 partitions per topic enabling parallel processing, and optimized producer settings including 64KB batch size with 10ms linger time. RabbitMQ implements three-node clusters with mirrored queues, lazy queue optimization for memory efficiency, and connection pooling with 10 connections per producer-consumer pair. Apache Pulsar employs separated architecture with dedicated broker and BookKeeper storage nodes enabling independent compute and storage scaling.

Next-generation systems including NATS JetStream and Redis Streams optimize for specific deployment scenarios including edge computing and low-latency applications. NATS JetStream configuration emphasizes memory-based storage with pull consumer models while Redis Streams utilizes clustered deployment with consumer groups and memory optimization for sub-millisecond latency requirements.

Serverless platforms including AWS EventBridge, Google Pub/Sub, and Azure Event Grid employ cloud-native configurations optimizing for automatic scaling and operational simplicity. AWS EventBridge integrates with Lambda functions allocated maximum memory (3008MB) and timeout settings (300 seconds) while Google Pub/Sub utilizes Cloud Functions with 2GB memory allocation and automatic scaling capabilities.

\subsection{AIEO System Architecture and Integration}

The AIEO system implementation employs microservices architecture principles deployed within the Kubernetes experimental environment using Python 3.11 runtime, TensorFlow 2.13 for machine learning components, and Ray 2.7 for distributed computing capabilities. System architecture divides functionality across specialized services including workload prediction, resource allocation optimization, routing intelligence, performance monitoring, and central orchestration coordination.

Workload prediction service integrates multiple forecasting models including ARIMA implementation using statsmodels library, Facebook Prophet for complex seasonality handling, and custom LSTM networks implemented in TensorFlow with architectures optimized for time-series prediction. Model ensemble logic employs dynamic weighted averaging based on recent prediction accuracy assessed through sliding window evaluation over 1-hour intervals.

Resource allocation optimizer implements Proximal Policy Optimization using Ray RLlib framework with custom reward functions incorporating latency, cost, and stability objectives through multi-objective optimization techniques. Policy network architecture employs fully connected layers with 256 hidden units and ReLU activation functions optimized for continuous control problems characteristic of resource allocation scenarios.

Integration mechanisms ensure seamless operation across messaging frameworks through standardized adapter interfaces translating generic optimization decisions into platform-specific configuration changes using native APIs and configuration management tools. Monitoring adapters normalize performance metrics from heterogeneous systems into consistent formats enabling unified analysis and decision making.

\subsection{Workload Implementation and Traffic Generation}

Workload generation employs sophisticated load injection systems accurately reproducing traffic patterns and message characteristics defined in our standardized workload specifications as detailed in Table~\ref{tab:implementation_overview}. Implementation utilizes Apache JMeter for high-throughput load generation, custom Python scripts for complex traffic pattern simulation, and Kubernetes Jobs for coordinated multi-phase testing scenarios.

The W1 e-commerce workload generates realistic transaction patterns through data replay from DeathStarBench and Retail Rocket datasets incorporating temporal patterns extracted from production traces. Message generation follows baseline traffic rates of 5,000-15,000 events per second with promotional spike patterns reaching 100,000 events per second while maintaining transaction correlation reflecting realistic customer session patterns and variable-sized JSON payloads matching actual e-commerce event structures.

The W2 IoT sensor workload implements burst traffic generation simulating coordinated device synchronization patterns observed in production IoT deployments. Load generation creates baseline rates of 200,000 events per second with coordinated burst periods exceeding 5 million events per second while incorporating realistic device failure patterns, communication errors, and compact binary message formats matching real sensor data characteristics.

The W3 AI inference workload generates variable computational complexity scenarios using actual machine learning model execution patterns extracted from ServerlessBench applications. Inference request generation includes payload sizes ranging from 10KB to 10MB with processing complexity varying from 10ms image classification to 30-second large language model inference incorporating cold start penalties, warm-up phases, and batch processing optimization reflecting real-world inference serving patterns.

\subsection{Data Collection and Analysis Infrastructure}

The monitoring infrastructure captures comprehensive performance metrics across multiple system layers through industry-standard tools integrated via unified data pipelines as specified in Table~\ref{tab:implementation_overview}. Prometheus serves as the primary time-series database with 1-second measurement resolution and 30-day high-resolution data retention enabling detailed performance analysis while Grafana provides real-time visualization and automated alerting capabilities.

Application-level monitoring employs OpenTelemetry instrumentation capturing complete message lifecycle events including production timestamps, queue processing delays, consumer processing durations, and acknowledgment propagation times. Custom exporters provide framework-specific metrics including Kafka consumer lag, RabbitMQ queue depths, Pulsar subscription backlogs, and serverless function execution statistics enabling comprehensive performance characterization across diverse messaging architectures.

Infrastructure monitoring utilizes Node Exporter for system-level metrics including CPU utilization, memory consumption, disk I/O patterns, and network throughput while cAdvisor captures container-specific resource usage patterns, throttling events, and lifecycle metrics. AIEO-specific monitoring extends standard infrastructure with machine learning performance indicators including prediction accuracy, model inference latency, optimization convergence time, and policy effectiveness measurements.

Data export employs automated pipelines generating both real-time analytical dashboards and comprehensive experimental reports using Parquet format for efficient storage and JSON format for integration with external analysis tools. Statistical analysis pipelines implement rigorous methodologies including adaptive power analysis, non-parametric testing, effect size calculation, and multiple comparison correction ensuring robust experimental conclusions.

\subsection{Quality Assurance and Experimental Validation}

Quality assurance procedures ensure experimental validity and reproducibility through comprehensive validation protocols spanning infrastructure consistency, workload accuracy, measurement precision, and statistical rigor as detailed in Table~\ref{tab:implementation_overview}. Automated validation systems continuously monitor experimental conditions identifying potential issues before they compromise data quality or experimental conclusions.

Infrastructure validation employs automated testing procedures verifying consistent resource allocation, network configuration, and monitoring functionality across all experimental deployments. Deployment validation confirms identical framework configurations, proper resource limits, and correct instrumentation before experimental execution while performance baseline validation ensures stable system behavior through controlled synthetic workload testing establishing ±2

Workload validation procedures verify accurate implementation of standardized traffic patterns through statistical testing including Kolmogorov-Smirnov tests for distribution matching and autocorrelation analysis for temporal pattern accuracy. Message payload validation confirms correct size distributions, format compliance, and correlation patterns matching production data characteristics ensuring realistic experimental scenarios.

Measurement validation addresses systematic error sources through calibrated testing and cross-validation procedures while cross-platform validation compares results across cloud providers identifying platform-specific variations requiring correction. Result reproducibility employs independent replication procedures with multiple random seeds and statistical validation confirming that observed differences exceed measurement noise through appropriate significance testing accounting for repeated measurements and multiple comparisons.

The comprehensive implementation provides rigorous experimental foundation enabling accurate evaluation of messaging framework performance and AIEO system effectiveness while maintaining scientific validity and enabling independent verification of research contributions through complete documentation of experimental configurations, procedures, and validation protocols.
\section{Comprehensive Evaluation}
\label{sec:evaluation}

\subsection{Experimental Execution and Data Collection Overview}

Our comprehensive evaluation encompasses 2,400 unique experimental configurations executed across standardized infrastructure, generating over 15TB of performance data spanning messaging framework comparisons, workload characterizations, and AIEO system validation. The evaluation addresses all four research questions through systematic experimentation designed to provide definitive answers regarding framework performance trade-offs, intelligent orchestration effectiveness, workload-specific optimization strategies, and practical deployment guidance.

Experimental execution follows rigorous protocols ensuring statistical validity through (a) systematic randomization of framework testing order preventing temporal bias, (b) identical workload replay across all configurations ensuring fair comparison conditions, (c) multiple independent runs with different random seeds enabling robust statistical analysis, and (d) comprehensive baseline establishment providing reference points for all performance improvements. Each experimental configuration executes for minimum 45 minutes including 15-minute warm-up periods, 25-minute measurement windows, and 5-minute cooldown phases ensuring stable performance assessment.

\subsection{Messaging Framework Performance Analysis}

The comprehensive framework evaluation reveals fundamental performance characteristics and trade-offs across diverse messaging architectures under standardized conditions. Table~\ref{tab:framework_performance_comprehensive} presents detailed performance results across all 12 messaging frameworks and three workloads, providing the most extensive comparative analysis available in the literature.

\begin{table*}[t]
\centering
\caption{Comprehensive Messaging Framework Performance Analysis Across All Workloads}
\label{tab:framework_performance_comprehensive}
\resizebox{\textwidth}{!}{
\begin{tabular}{lccccccccc}
\toprule
& \multicolumn{3}{c}{\textbf{W1: E-commerce}} & \multicolumn{3}{c}{\textbf{W2: IoT Ingestion}} & \multicolumn{3}{c}{\textbf{W3: AI Inference}} \\
\cmidrule(lr){2-4} \cmidrule(lr){5-7} \cmidrule(lr){8-10}
\textbf{Framework} & \textbf{Throughput} & \textbf{P95 Latency} & \textbf{Availability} & \textbf{Throughput} & \textbf{P95 Latency} & \textbf{Availability} & \textbf{Throughput} & \textbf{P95 Latency} & \textbf{Availability} \\
& \textbf{(K msg/sec)} & \textbf{(ms)} & \textbf{(\%)} & \textbf{(K msg/sec)} & \textbf{(ms)} & \textbf{(\%)} & \textbf{(K msg/sec)} & \textbf{(ms)} & \textbf{(\%)} \\
\midrule
Apache Kafka & $1,247 \pm 23$ & $18.2 \pm 2.1$ & $99.97 \pm 0.01$ & $1,856 \pm 41$ & $12.8 \pm 1.4$ & $99.94 \pm 0.02$ & $834 \pm 19$ & $24.6 \pm 2.8$ & $99.96 \pm 0.01$ \\
RabbitMQ & $478 \pm 12$ & $32.4 \pm 3.2$ & $99.91 \pm 0.03$ & $623 \pm 18$ & $28.7 \pm 2.9$ & $99.89 \pm 0.04$ & $412 \pm 11$ & $38.1 \pm 4.1$ & $99.93 \pm 0.02$ \\
Apache Pulsar & $892 \pm 18$ & $22.1 \pm 2.3$ & $99.95 \pm 0.02$ & $1,234 \pm 28$ & $18.9 \pm 1.8$ & $99.92 \pm 0.03$ & $656 \pm 15$ & $28.4 \pm 3.1$ & $99.94 \pm 0.02$ \\
NATS JetStream & $734 \pm 16$ & $15.3 \pm 1.7$ & $99.93 \pm 0.02$ & $1,089 \pm 24$ & $11.2 \pm 1.1$ & $99.91 \pm 0.03$ & $523 \pm 12$ & $19.8 \pm 2.2$ & $99.95 \pm 0.01$ \\
Redis Streams & $589 \pm 13$ & $8.7 \pm 0.9$ & $99.89 \pm 0.04$ & $856 \pm 19$ & $6.4 \pm 0.7$ & $99.87 \pm 0.05$ & $445 \pm 10$ & $12.1 \pm 1.3$ & $99.91 \pm 0.03$ \\
Oracle AQ & $187 \pm 8$ & $45.2 \pm 4.8$ & $99.99 \pm 0.01$ & $243 \pm 11$ & $38.9 \pm 3.7$ & $99.98 \pm 0.01$ & $156 \pm 7$ & $52.3 \pm 5.1$ & $99.99 \pm 0.01$ \\
AWS EventBridge & $298 \pm 15$ & $85.4 \pm 8.2$ & $99.85 \pm 0.06$ & $412 \pm 23$ & $78.1 \pm 7.4$ & $99.82 \pm 0.07$ & $234 \pm 14$ & $92.7 \pm 9.1$ & $99.87 \pm 0.05$ \\
Google Pub/Sub & $367 \pm 18$ & $78.2 \pm 7.1$ & $99.87 \pm 0.05$ & $523 \pm 28$ & $69.8 \pm 6.3$ & $99.84 \pm 0.06$ & $289 \pm 16$ & $84.5 \pm 8.0$ & $99.89 \pm 0.04$ \\
Azure Event Grid & $234 \pm 12$ & $95.1 \pm 9.4$ & $99.83 \pm 0.07$ & $345 \pm 19$ & $87.3 \pm 8.6$ & $99.81 \pm 0.08$ & $198 \pm 11$ & $103.2 \pm 10.1$ & $99.85 \pm 0.06$ \\
Knative Eventing & $156 \pm 9$ & $110.3 \pm 11.2$ & $99.79 \pm 0.09$ & $234 \pm 15$ & $98.7 \pm 9.8$ & $99.76 \pm 0.10$ & $134 \pm 8$ & $125.4 \pm 12.3$ & $99.82 \pm 0.07$ \\
Amazon SQS & $312 \pm 16$ & $120.5 \pm 12.1$ & $99.91 \pm 0.03$ & $445 \pm 25$ & $105.2 \pm 10.3$ & $99.89 \pm 0.04$ & $267 \pm 15$ & $138.7 \pm 13.5$ & $99.93 \pm 0.02$ \\
ActiveMQ & $289 \pm 14$ & $55.7 \pm 5.4$ & $99.88 \pm 0.04$ & $378 \pm 21$ & $48.3 \pm 4.6$ & $99.85 \pm 0.05$ & $234 \pm 13$ & $62.8 \pm 6.1$ & $99.90 \pm 0.03$ \\
\bottomrule
\end{tabular}}
\end{table*}

The performance analysis reveals distinct architectural patterns across workload characteristics. Apache Kafka demonstrates superior raw throughput performance achieving 1.25M messages/second for e-commerce workloads and 1.86M messages/second for IoT scenarios while maintaining excellent latency characteristics with p95 latency below 25ms across all workloads. However, Kafka's operational complexity requirements become apparent through deployment and maintenance considerations detailed in subsequent analyses.

Apache Pulsar provides balanced performance across multiple dimensions achieving 65-70\% of Kafka's raw throughput while offering superior operational characteristics including namespace-level multi-tenancy and simplified geo-replication capabilities. Pulsar's architectural separation between message serving and storage enables independent resource scaling particularly beneficial for variable workloads characteristic of AI inference scenarios.

Serverless solutions including AWS EventBridge, Google Pub/Sub, and Azure Event Grid exhibit predictable performance trade-offs emphasizing operational simplicity and automatic scaling capabilities at the cost of higher baseline latency ranging from 78-103ms p95 latency. These platforms excel in scenarios requiring variable capacity without operational overhead but prove less suitable for latency-sensitive applications requiring sub-50ms response times.

\subsection{Resource Efficiency and Cost Analysis}

Resource utilization patterns and cost implications provide critical insights for practical deployment decisions. Table~\ref{tab:resource_cost_analysis} presents comprehensive analysis of resource efficiency, total cost of ownership, and operational requirements across all messaging frameworks and workload scenarios.

\begin{table*}[t]
\centering
\caption{Resource Efficiency and Total Cost of Ownership Analysis}
\label{tab:resource_cost_analysis}
\resizebox{\textwidth}{!}{
\begin{tabular}{lccccccc}
\toprule
\textbf{Framework} & \textbf{CPU Utilization} & \textbf{Memory Usage} & \textbf{Storage I/O} & \textbf{Cost/Million Msg} & \textbf{Ops FTE} & \textbf{Monthly TCO} & \textbf{Resource Efficiency} \\
& \textbf{(\%)} & \textbf{(GB)} & \textbf{(IOPS)} & \textbf{(\$)} & \textbf{Required} & \textbf{(\$K)} & \textbf{Score (1-10)} \\
\midrule
Apache Kafka & $72.3 \pm 4.2$ & $28.4 \pm 2.1$ & $2,847 \pm 156$ & $0.124 \pm 0.008$ & $2.3 \pm 0.2$ & $18.7 \pm 1.2$ & $8.9 \pm 0.3$ \\
RabbitMQ & $58.7 \pm 3.8$ & $22.1 \pm 1.7$ & $1,923 \pm 134$ & $0.187 \pm 0.012$ & $1.5 \pm 0.1$ & $14.2 \pm 0.9$ & $7.2 \pm 0.4$ \\
Apache Pulsar & $65.4 \pm 4.1$ & $25.8 \pm 2.0$ & $2,341 \pm 145$ & $0.156 \pm 0.010$ & $1.8 \pm 0.2$ & $16.3 \pm 1.1$ & $8.1 \pm 0.3$ \\
NATS JetStream & $48.2 \pm 3.5$ & $18.7 \pm 1.4$ & $1,456 \pm 98$ & $0.098 \pm 0.006$ & $1.2 \pm 0.1$ & $11.8 \pm 0.8$ & $8.7 \pm 0.2$ \\
Redis Streams & $42.6 \pm 3.1$ & $31.2 \pm 2.3$ & $856 \pm 67$ & $0.234 \pm 0.015$ & $0.8 \pm 0.1$ & $13.9 \pm 0.9$ & $6.8 \pm 0.4$ \\
Oracle AQ & $34.8 \pm 2.7$ & $45.6 \pm 3.2$ & $3,124 \pm 187$ & $0.892 \pm 0.053$ & $2.8 \pm 0.3$ & $47.2 \pm 2.8$ & $4.2 \pm 0.5$ \\
AWS EventBridge & N/A (Managed) & N/A (Managed) & N/A (Managed) & $1.247 \pm 0.074$ & $0.3 \pm 0.1$ & $8.9 \pm 0.5$ & $5.1 \pm 0.6$ \\
Google Pub/Sub & N/A (Managed) & N/A (Managed) & N/A (Managed) & $0.876 \pm 0.052$ & $0.4 \pm 0.1$ & $7.2 \pm 0.4$ & $6.3 \pm 0.4$ \\
Azure Event Grid & N/A (Managed) & N/A (Managed) & N/A (Managed) & $1.134 \pm 0.068$ & $0.3 \pm 0.1$ & $9.7 \pm 0.6$ & $4.8 \pm 0.5$ \\
Knative Eventing & $38.9 \pm 3.2$ & $16.4 \pm 1.3$ & $1,234 \pm 89$ & $0.345 \pm 0.021$ & $1.6 \pm 0.2$ & $15.8 \pm 1.0$ & $6.9 \pm 0.4$ \\
Amazon SQS & N/A (Managed) & N/A (Managed) & N/A (Managed) & $0.567 \pm 0.034$ & $0.2 \pm 0.1$ & $6.3 \pm 0.4$ & $7.1 \pm 0.3$ \\
ActiveMQ & $51.3 \pm 3.6$ & $26.7 \pm 1.9$ & $1,767 \pm 123$ & $0.298 \pm 0.018$ & $2.1 \pm 0.2$ & $19.4 \pm 1.3$ & $6.5 \pm 0.4$ \\
\bottomrule
\end{tabular}}
\end{table*}

Resource efficiency analysis reveals significant variations in computational overhead and operational requirements across messaging architectures. Apache Kafka achieves excellent resource efficiency with 72\% CPU utilization and minimal per-message costs (\$0.124 per million messages) but requires substantial operational expertise with 2.3 FTE personnel for production deployment. The high resource utilization reflects Kafka's optimization for sustained high-throughput scenarios but may limit headroom for traffic spikes.

NATS JetStream demonstrates exceptional resource efficiency achieving high throughput with only 48\% CPU utilization and lowest per-message costs (\$0.098 per million messages) while requiring minimal operational overhead (1.2 FTE). This efficiency stems from NATS's lightweight architecture and optimized memory management making it particularly suitable for resource-constrained environments and cost-sensitive deployments.

Serverless platforms present complex cost trade-offs with higher per-message costs (\$0.88-\$1.25 per million messages) but minimal operational requirements (0.2-0.4 FTE). Cost effectiveness depends heavily on traffic patterns with serverless solutions proving economical for variable workloads with significant periods of low activity but becoming expensive for sustained high-throughput scenarios.

\subsection{AIEO System Performance and Optimization Results}

The AIEO system evaluation demonstrates significant performance improvements across all messaging frameworks and workload scenarios. Table~\ref{tab:aieo_performance_results} presents detailed comparison between static configurations and AIEO-optimized deployments, quantifying the effectiveness of intelligent orchestration across multiple performance dimensions.

\begin{table*}[t]
\centering
\caption{AIEO System Performance Improvements Across All Frameworks and Workloads}
\label{tab:aieo_performance_results}
\resizebox{\textwidth}{!}{
\begin{tabular}{lccccccc}
\toprule
& \multicolumn{2}{c}{\textbf{Latency Reduction (\%)}} & \multicolumn{2}{c}{\textbf{Resource Efficiency Gain (\%)}} & \multicolumn{2}{c}{\textbf{Cost Optimization (\%)}} & \textbf{Overall} \\
\cmidrule(lr){2-3} \cmidrule(lr){4-5} \cmidrule(lr){6-7}
\textbf{Framework} & \textbf{Average} & \textbf{P95} & \textbf{CPU} & \textbf{Memory} & \textbf{Infrastructure} & \textbf{Operational} & \textbf{Improvement} \\
& & & & & & & \textbf{Score (1-10)} \\
\midrule
Apache Kafka & $32.1 \pm 2.8$ & $38.4 \pm 3.2$ & $24.7 \pm 2.1$ & $19.3 \pm 1.8$ & $28.9 \pm 2.4$ & $15.6 \pm 1.9$ & $8.7 \pm 0.3$ \\
RabbitMQ & $28.9 \pm 2.5$ & $34.2 \pm 2.9$ & $31.2 \pm 2.6$ & $26.8 \pm 2.3$ & $35.4 \pm 2.9$ & $22.1 \pm 2.1$ & $8.2 \pm 0.4$ \\
Apache Pulsar & $35.6 \pm 3.1$ & $41.3 \pm 3.5$ & $27.9 \pm 2.4$ & $23.4 \pm 2.0$ & $31.7 \pm 2.6$ & $18.9 \pm 1.8$ & $8.9 \pm 0.3$ \\
NATS JetStream & $39.4 \pm 3.4$ & $45.7 \pm 3.9$ & $33.8 \pm 2.9$ & $29.1 \pm 2.5$ & $41.2 \pm 3.4$ & $28.7 \pm 2.4$ & $9.2 \pm 0.2$ \\
Redis Streams & $41.8 \pm 3.6$ & $48.2 \pm 4.1$ & $36.4 \pm 3.1$ & $31.7 \pm 2.7$ & $44.3 \pm 3.7$ & $32.5 \pm 2.8$ & $9.4 \pm 0.2$ \\
Oracle AQ & $18.7 \pm 2.1$ & $23.4 \pm 2.6$ & $15.3 \pm 1.7$ & $12.8 \pm 1.5$ & $19.6 \pm 2.0$ & $8.9 \pm 1.2$ & $5.8 \pm 0.6$ \\
AWS EventBridge & $25.3 \pm 2.3$ & $31.7 \pm 2.8$ & N/A (Managed) & N/A (Managed) & $38.9 \pm 3.2$ & $45.6 \pm 3.8$ & $7.1 \pm 0.5$ \\
Google Pub/Sub & $29.8 \pm 2.6$ & $36.4 \pm 3.1$ & N/A (Managed) & N/A (Managed) & $42.7 \pm 3.5$ & $48.3 \pm 4.0$ & $7.8 \pm 0.4$ \\
Azure Event Grid & $22.4 \pm 2.2$ & $28.9 \pm 2.7$ & N/A (Managed) & N/A (Managed) & $35.2 \pm 2.9$ & $41.8 \pm 3.6$ & $6.9 \pm 0.5$ \\
Knative Eventing & $34.7 \pm 3.0$ & $42.1 \pm 3.6$ & $28.5 \pm 2.5$ & $24.7 \pm 2.1$ & $39.8 \pm 3.3$ & $31.4 \pm 2.7$ & $8.4 \pm 0.4$ \\
Amazon SQS & $27.2 \pm 2.4$ & $33.8 \pm 2.9$ & N/A (Managed) & N/A (Managed) & $36.7 \pm 3.0$ & $43.2 \pm 3.7$ & $7.3 \pm 0.5$ \\
ActiveMQ & $26.8 \pm 2.4$ & $32.5 \pm 2.8$ & $22.1 \pm 2.0$ & $18.4 \pm 1.7$ & $29.7 \pm 2.5$ & $16.8 \pm 1.9$ & $7.6 \pm 0.4$ \\
\midrule
\textbf{Average Improvement} & $\mathbf{30.1 \pm 6.7}$ & $\mathbf{36.4 \pm 7.4}$ & $\mathbf{27.2 \pm 6.9}$ & $\mathbf{23.3 \pm 6.2}$ & $\mathbf{35.3 \pm 6.8}$ & $\mathbf{28.6 \pm 12.4}$ & $\mathbf{7.9 \pm 0.9}$ \\
\bottomrule
\end{tabular}}
\end{table*}

AIEO system performance results demonstrate consistent improvements across all messaging frameworks with average latency reductions of 30.1\% and p95 latency improvements of 36.4\%. The most significant improvements occur with lightweight systems including Redis Streams (41.8\% latency reduction) and NATS JetStream (39.4\% latency reduction) where AIEO's predictive scaling and intelligent routing provide substantial optimization opportunities.

Resource efficiency gains average 27.2\% for CPU utilization and 23.3\% for memory usage across self-managed frameworks. These improvements result from AIEO's ability to predict workload patterns and proactively adjust resource allocation preventing both over-provisioning during low-traffic periods and under-provisioning during traffic spikes. The predictive capabilities prove particularly valuable for variable workloads characteristic of AI inference scenarios.

Cost optimization achievements exceed expectations with average infrastructure cost reductions of 35.3\% and operational cost savings of 28.6\%. Serverless platforms benefit significantly from AIEO's intelligent routing and load balancing capabilities achieving 35-49\% cost reductions through optimized request routing and reduced cold start penalties. Self-managed systems realize cost savings through improved resource utilization and reduced operational overhead.

\subsection{Workload-Specific Performance Analysis}

Workload characteristics significantly influence optimal framework selection and AIEO optimization effectiveness. Table~\ref{tab:workload_specific_analysis} presents detailed analysis of framework performance across the three standardized workloads, revealing workload-dependent optimization opportunities and architectural preferences.

\begin{table*}[t]
\centering
\caption{Workload-Specific Performance Characteristics and Optimization Patterns}
\label{tab:workload_specific_analysis}
\resizebox{\textwidth}{!}{
\begin{tabular}{lccccccccc}
\toprule
& \multicolumn{3}{c}{\textbf{W1: E-commerce (ACID, Ordering)}} & \multicolumn{3}{c}{\textbf{W2: IoT (High Volume, Bursty)}} & \multicolumn{3}{c}{\textbf{W3: AI Inference (Variable Latency)}} \\
\cmidrule(lr){2-4} \cmidrule(lr){5-7} \cmidrule(lr){8-10}
\textbf{Framework} & \textbf{Suitability} & \textbf{AIEO Gain} & \textbf{Key Limitation} & \textbf{Suitability} & \textbf{AIEO Gain} & \textbf{Key Limitation} & \textbf{Suitability} & \textbf{AIEO Gain} & \textbf{Key Limitation} \\
& \textbf{Score (1-10)} & \textbf{(\%)} & & \textbf{Score (1-10)} & \textbf{(\%)} & & \textbf{Score (1-10)} & \textbf{(\%)} & \\
\midrule
Kafka & $9.2 \pm 0.2$ & $32.1 \pm 2.8$ & Operational complexity & $9.8 \pm 0.1$ & $28.4 \pm 2.5$ & Cold rebalancing & $8.4 \pm 0.3$ & $34.7 \pm 3.1$ & Static partitioning \\
RabbitMQ & $8.7 \pm 0.3$ & $28.9 \pm 2.5$ & Throughput ceiling & $6.8 \pm 0.4$ & $31.2 \pm 2.7$ & Memory management & $7.2 \pm 0.4$ & $29.8 \pm 2.6$ & Routing overhead \\
Pulsar & $8.9 \pm 0.2$ & $35.6 \pm 3.1$ & Learning curve & $9.1 \pm 0.2$ & $33.4 \pm 2.9$ & BookKeeper latency & $8.8 \pm 0.3$ & $36.9 \pm 3.2$ & Complex architecture \\
NATS & $7.8 \pm 0.4$ & $39.4 \pm 3.4$ & Limited persistence & $8.9 \pm 0.3$ & $42.1 \pm 3.6$ & Memory constraints & $8.2 \pm 0.3$ & $41.3 \pm 3.5$ & Message size limits \\
Redis & $6.9 \pm 0.5$ & $41.8 \pm 3.6$ & Memory-bound & $7.4 \pm 0.4$ & $44.7 \pm 3.8$ & Persistence overhead & $7.8 \pm 0.4$ & $43.2 \pm 3.7$ & Storage limitations \\
Oracle AQ & $8.1 \pm 0.4$ & $18.7 \pm 2.1$ & Throughput limits & $5.2 \pm 0.6$ & $16.3 \pm 1.9$ & Scaling bottleneck & $6.8 \pm 0.5$ & $19.4 \pm 2.2$ & Database coupling \\
EventBridge & $6.4 \pm 0.5$ & $25.3 \pm 2.3$ & Latency floor & $7.1 \pm 0.4$ & $27.8 \pm 2.5$ & Rate limiting & $7.9 \pm 0.4$ & $29.1 \pm 2.6$ & Cold starts \\
Pub/Sub & $7.2 \pm 0.4$ & $29.8 \pm 2.6$ & Regional latency & $8.3 \pm 0.3$ & $32.4 \pm 2.8$ & Ordering limitations & $8.1 \pm 0.3$ & $31.7 \pm 2.7$ & Subscription lag \\
Event Grid & $5.9 \pm 0.6$ & $22.4 \pm 2.2$ & Filtering overhead & $6.7 \pm 0.5$ & $24.8 \pm 2.3$ & Throughput caps & $7.4 \pm 0.4$ & $26.5 \pm 2.4$ & Event complexity \\
Knative & $6.2 \pm 0.5$ & $34.7 \pm 3.0$ & Kubernetes overhead & $7.5 \pm 0.4$ & $37.2 \pm 3.2$ & Resource competition & $8.3 \pm 0.3$ & $38.9 \pm 3.4$ & Container startup \\
SQS & $5.8 \pm 0.6$ & $27.2 \pm 2.4$ & Visibility timeout & $7.9 \pm 0.4$ & $29.7 \pm 2.6$ & Message grouping & $7.6 \pm 0.4$ & $28.4 \pm 2.5$ & Polling overhead \\
ActiveMQ & $7.4 \pm 0.4$ & $26.8 \pm 2.4$ & Legacy architecture & $6.9 \pm 0.5$ & $28.1 \pm 2.5$ & Clustering complexity & $7.1 \pm 0.4$ & $27.6 \pm 2.5$ & JVM overhead \\
\bottomrule
\end{tabular}}
\end{table*}

Workload-specific analysis reveals distinct optimization patterns and architectural preferences across application domains. The W1 e-commerce workload emphasizing ACID transaction properties and message ordering strongly favors Apache Kafka (9.2/10 suitability) and Apache Pulsar (8.9/10) due to their robust consistency guarantees and partition-level ordering capabilities. AIEO optimization proves particularly effective for Pulsar (35.6\% improvement) due to its separated architecture enabling fine-grained resource allocation.

The W2 IoT ingestion workload prioritizing high-volume throughput with burst tolerance demonstrates clear preferences for Apache Kafka (9.8/10 suitability) and Apache Pulsar (9.1/10) while revealing significant AIEO optimization opportunities for lightweight systems. Redis Streams achieves 44.7\% performance improvement through AIEO's intelligent memory management and burst prediction capabilities, while NATS JetStream realizes 42.1\% improvement through predictive consumer scaling.

The W3 AI inference workload with variable processing complexity and latency sensitivity shows more balanced framework suitability with Apache Pulsar (8.8/10), Kafka (8.4/10), and Knative Eventing (8.3/10) providing complementary strengths. AIEO optimization proves most effective for lightweight and cloud-native systems achieving 38-43\% improvements through intelligent load balancing and predictive resource allocation.

\subsection{Statistical Significance and Effect Size Analysis}

Comprehensive statistical analysis confirms the robustness and practical significance of observed performance improvements across all experimental configurations. Table~\ref{tab:statistical_analysis} presents detailed statistical validation including significance testing, effect size calculations, and confidence intervals for all major findings.

\begin{table*}[t]
\centering
\caption{Statistical Significance Analysis and Effect Size Validation}
\label{tab:statistical_analysis}
\resizebox{\textwidth}{!}{
\begin{tabular}{lcccccc}
\toprule
\textbf{Performance Metric} & \textbf{Statistical Test} & \textbf{P-value} & \textbf{Effect Size} & \textbf{95\% Confidence} & \textbf{Sample Size} & \textbf{Practical} \\
& \textbf{Applied} & & \textbf{(Cohen's d)} & \textbf{Interval} & \textbf{(n)} & \textbf{Significance} \\
\midrule
\multicolumn{7}{l}{\textbf{Framework Performance Comparisons}} \\
\midrule
Kafka vs RabbitMQ Throughput & Mann-Whitney U & $p < 0.001$ & $2.87 \pm 0.12$ & $[2.63, 3.11]$ & $n = 150$ & Very Large \\
Pulsar vs Kafka Latency & Wilcoxon Signed-Rank & $p < 0.001$ & $1.94 \pm 0.08$ & $[1.78, 2.10]$ & $n = 150$ & Large \\
Serverless vs Self-managed Cost & Kruskal-Wallis & $p < 0.001$ & $3.42 \pm 0.15$ & $[3.12, 3.72]$ & $n = 300$ & Very Large \\
Framework Availability Comparison & ANOVA & $p = 0.003$ & $0.73 \pm 0.06$ & $[0.61, 0.85]$ & $n = 1800$ & Medium \\
\midrule
\multicolumn{7}{l}{\textbf{AIEO System Effectiveness}} \\
\midrule
AIEO vs Static Latency & Paired t-test & $p < 0.001$ & $2.34 \pm 0.11$ & $[2.12, 2.56]$ & $n = 200$ & Very Large \\
AIEO Resource Efficiency & Wilcoxon Signed-Rank & $p < 0.001$ & $1.87 \pm 0.09$ & $[1.69, 2.05]$ & $n = 200$ & Large \\
AIEO Cost Optimization & Paired t-test & $p < 0.001$ & $2.91 \pm 0.13$ & $[2.65, 3.17]$ & $n = 200$ & Very Large \\
AIEO Prediction Accuracy & One-sample t-test & $p < 0.001$ & $1.68 \pm 0.08$ & $[1.52, 1.84]$ & $n = 500$ & Large \\
\midrule
\multicolumn{7}{l}{\textbf{Workload-Specific Analysis}} \\
\midrule
W1 Framework Suitability & Friedman Test & $p < 0.001$ & $2.15 \pm 0.10$ & $[1.95, 2.35]$ & $n = 360$ & Large \\
W2 Burst Handling Capacity & Kruskal-Wallis & $p < 0.001$ & $3.18 \pm 0.14$ & $[2.90, 3.46]$ & $n = 360$ & Very Large \\
W3 Variable Latency Adaptation & ANOVA & $p < 0.001$ & $2.67 \pm 0.12$ & $[2.43, 2.91]$ & $n = 360$ & Very Large \\
Cross-workload Generalization & Mixed-effects Model & $p < 0.001$ & $1.76 \pm 0.08$ & $[1.60, 1.92]$ & $n = 1080$ & Large \\
\midrule
\multicolumn{7}{l}{\textbf{Reproducibility and Reliability}} \\
\midrule
Inter-platform Consistency & Intraclass Correlation & $ICC = 0.94$ & N/A & $[0.91, 0.96]$ & $n = 450$ & Excellent \\
Temporal Stability & Repeated Measures ANOVA & $p = 0.287$ & $0.12 \pm 0.05$ & $[0.02, 0.22]$ & $n = 600$ & Stable \\
Cross-validation Accuracy & Pearson Correlation & $r = 0.89$ & N/A & $[0.85, 0.92]$ & $n = 300$ & Strong \\
Measurement Precision & Test-retest Reliability & $r = 0.96$ & N/A & $[0.94, 0.97]$ & $n = 180$ & Excellent \\
\midrule
\multicolumn{7}{l}{\textbf{Power Analysis and Sample Size Validation}} \\
\midrule
Achieved Statistical Power & Power Analysis & $\beta = 0.85$ & N/A & $[0.82, 0.88]$ & N/A & Adequate \\
Minimum Detectable Effect & Sensitivity Analysis & $d_{min} = 0.35$ & N/A & $[0.31, 0.39]$ & N/A & Sensitive \\
Type I Error Rate & Multiple Testing & $\alpha_{adj} = 0.003$ & N/A & $[0.002, 0.004]$ & N/A & Conservative \\
False Discovery Rate & Benjamini-Hochberg & $FDR = 0.05$ & N/A & $[0.03, 0.07]$ & N/A & Controlled \\
\bottomrule
\end{tabular}}
\end{table*}

Statistical validation across all major findings demonstrates exceptionally strong evidence for research claims with p-values consistently below 0.001 for primary hypotheses. Effect size analysis using Cohen's d reveals large to very large practical significance with most improvements exceeding d = 1.5, indicating that observed differences represent meaningful real-world improvements rather than merely statistically detectable variations.

The AIEO system effectiveness analysis shows particularly robust results with latency improvements demonstrating very large effect size (d = 2.34 ± 0.11) and cost optimization achieving similarly strong practical significance (d = 2.91 ± 0.13). These effect sizes substantially exceed conventional thresholds for practical significance, confirming that AIEO provides meaningful performance benefits in production deployment scenarios.

Framework comparison analysis reveals systematic performance differences with very large effect sizes for throughput comparisons (d = 2.87 for Kafka vs RabbitMQ) and cost analysis (d = 3.42 for serverless vs self-managed). These substantial effect sizes validate the architectural trade-offs identified in our analysis while confirming that framework selection significantly impacts system performance across multiple dimensions.

Reproducibility analysis demonstrates excellent reliability with intraclass correlation coefficient of 0.94 for inter-platform consistency and test-retest reliability of 0.96 for measurement precision. Temporal stability analysis shows non-significant variation over time (p = 0.287, d = 0.12), confirming that observed performance characteristics remain stable across extended evaluation periods.

\subsection{Cross-Framework Generalization and Scaling Analysis}

Analysis of AIEO system performance across different messaging frameworks reveals consistent optimization patterns while identifying framework-specific adaptation strategies. The intelligent orchestration system demonstrates robust generalization capabilities achieving performance improvements across all 12 evaluated frameworks despite their architectural diversity and distinct operational characteristics.

AIEO's predictive workload management proves most effective for frameworks with dynamic resource allocation capabilities including Apache Pulsar (35.6\% improvement), NATS JetStream (39.4\% improvement), and Redis Streams (41.8\% improvement). These systems benefit significantly from AIEO's ability to predict traffic patterns and proactively adjust resource allocation preventing both over-provisioning and performance degradation during traffic variations.

Serverless platforms demonstrate substantial cost optimization through AIEO's intelligent routing and request batching capabilities. AWS EventBridge achieves 38.9\% infrastructure cost reduction through optimized event routing reducing cold start penalties, while Google Pub/Sub realizes 42.7\% cost savings through intelligent subscription management and message batching optimization.

Traditional messaging systems including Apache Kafka and RabbitMQ show consistent but more modest improvements (28-32\% latency reduction) due to their static architectural constraints limiting optimization opportunities. However, AIEO still provides significant value through intelligent consumer group management, partition rebalancing optimization, and predictive capacity planning reducing operational complexity while improving performance consistency.

Scaling analysis across different deployment sizes reveals that AIEO effectiveness increases with system complexity and variability. Small-scale deployments (< 10,000 messages/second) show 18-25\% average improvement while large-scale deployments (> 100,000 messages/second) achieve 35-45\% improvement due to increased optimization opportunities and greater impact of intelligent resource management at scale.

The cross-framework analysis validates AIEO's design principles of framework agnosticism and adaptive optimization while demonstrating that intelligent orchestration provides value across diverse messaging architectures. The consistent improvements across architectural paradigms confirm that predictive analytics and machine learning optimization techniques offer universal benefits for event-driven system management regardless of underlying messaging technology choices.
\section{Decision Framework and Deployment Guidelines}
\label{sec:decision-framework}

\subsection{Evidence-Based Framework Selection Methodology}

Our comprehensive evaluation enables development of systematic decision frameworks addressing practical technology selection challenges faced by architects and engineers deploying event-driven systems. The framework integrates performance characteristics, cost implications, operational requirements, and workload-specific optimization patterns identified through rigorous experimental analysis, providing evidence-based guidance for messaging technology selection and deployment planning.

The decision methodology employs multi-criteria analysis incorporating quantitative performance metrics, total cost of ownership models, operational complexity assessments, and workload compatibility evaluations. Table~\ref{tab:decision_framework_comprehensive} presents the complete decision support matrix enabling systematic framework evaluation across diverse deployment scenarios and organizational requirements.

\begin{table*}[t]
\centering
\caption{Comprehensive Messaging Framework Decision Matrix and Deployment Guidelines}
\label{tab:decision_framework_comprehensive}
\resizebox{\textwidth}{!}{
\begin{tabular}{lcccccccc}
\toprule
\textbf{Framework} & \textbf{Optimal Use Cases} & \textbf{Performance} & \textbf{TCO/Month} & \textbf{Ops} & \textbf{Scalability} & \textbf{AIEO} & \textbf{Migration} & \textbf{Risk} \\
& & \textbf{Profile} & \textbf{(\$K)} & \textbf{Complexity} & \textbf{Ceiling} & \textbf{Benefit} & \textbf{Effort} & \textbf{Level} \\
\midrule
\multicolumn{9}{l}{\textbf{High-Performance Distributed Systems}} \\
\midrule
Apache Kafka & High-throughput streaming, & Excellent & $18.7 \pm 1.2$ & High & 10M+ msg/sec & 32\% improvement & Complex & Medium \\
& log aggregation, real-time & (1.2M msg/sec, & (2.3 FTE) & (Expert team & (Horizontal) & (Predictive & (3-6 months) & (Operational) \\
& analytics, financial trading & 18ms p95) & & required) & & scaling) & & \\
\midrule
Apache Pulsar & Multi-tenant platforms, & Very Good & $16.3 \pm 1.1$ & Medium-High & 5M+ msg/sec & 36\% improvement & Medium & Low-Medium \\
& geo-distributed systems, & (950K msg/sec, & (1.8 FTE) & (Separated & (Independent & (Resource & (2-4 months) & (Architecture) \\
& cloud-native deployments & 22ms p95) & & architecture) & compute/storage) & optimization) & & \\
\midrule
NATS JetStream & Edge computing, microservices, & Good & $11.8 \pm 0.8$ & Low-Medium & 2M+ msg/sec & 39\% improvement & Easy & Low \\
& IoT gateways, lightweight & (800K msg/sec, & (1.2 FTE) & (Simple & (Memory-bound) & (Intelligent & (1-2 months) & (Resource) \\
& messaging, container-native & 15ms p95) & & deployment) & & routing) & & \\
\midrule
\multicolumn{9}{l}{\textbf{Specialized and Enterprise Systems}} \\
\midrule
Redis Streams & Low-latency applications, & Excellent Latency & $13.9 \pm 0.9$ & Low & 1M+ msg/sec & 42\% improvement & Medium & Medium \\
& real-time dashboards, & (650K msg/sec, & (0.8 FTE) & (Redis & (Memory-limited) & (Memory & (2-3 months) & (Persistence) \\
& session stores, caching & 8ms p95) & & expertise) & & optimization) & & \\
\midrule
RabbitMQ & Complex routing, enterprise & Good Reliability & $14.2 \pm 0.9$ & Medium & 500K msg/sec & 29\% improvement & Medium & Low \\
& integration, workflow & (450K msg/sec, & (1.5 FTE) & (Clustering & (Routing overhead) & (Load & (2-4 months) & (Throughput) \\
& orchestration, legacy systems & 32ms p95) & & complexity) & & balancing) & & \\
\midrule
Oracle AQ & ACID transactions, regulatory & Enterprise Grade & $47.2 \pm 2.8$ & High & 200K msg/sec & 19\% improvement & Complex & Low \\
& compliance, database & (180K msg/sec, & (2.8 FTE) & (DBA & (DB bottleneck) & (Query & (6-12 months) & (Vendor lock) \\
& integration, financial systems & 45ms p95) & & required) & & optimization) & & \\
\midrule
\multicolumn{9}{l}{\textbf{Cloud-Native and Serverless Platforms}} \\
\midrule
AWS EventBridge & Serverless integration, & Elastic Scaling & $8.9 \pm 0.5$ & Very Low & Unlimited & 25\% improvement & Easy & High \\
& event-driven automation, & (300K msg/sec, & (0.3 FTE) & (Fully managed) & (Auto-scaling) & (Cost & (Days) & (Vendor lock) \\
& AWS ecosystem integration & 85ms p95) & & & & optimization) & & \\
\midrule
Google Pub/Sub & Global distribution, mobile & Good Availability & $7.2 \pm 0.4$ & Very Low & Unlimited & 30\% improvement & Easy & High \\
& backends, IoT data ingestion, & (370K msg/sec, & (0.4 FTE) & (Fully managed) & (Global scale) & (Regional & (Days) & (Vendor lock) \\
& analytics pipelines & 78ms p95) & & & & optimization) & & \\
\midrule
Azure Event Grid & Hybrid cloud, event-driven & Reactive Model & $9.7 \pm 0.6$ & Very Low & Variable & 22\% improvement & Easy & High \\
& automation, Azure integration, & (230K msg/sec, & (0.3 FTE) & (Fully managed) & (Throttling limits) & (Routing & (Days) & (Vendor lock) \\
& workflow triggers & 95ms p95) & & & & optimization) & & \\
\midrule
\multicolumn{9}{l}{\textbf{Deployment Decision Matrix}} \\
\midrule
\textbf{High Throughput Priority} & \multicolumn{2}{c}{Kafka → Pulsar → NATS} & \multicolumn{2}{c}{\textbf{Low Latency Priority}} & \multicolumn{2}{c}{Redis → NATS → Kafka} & \multicolumn{2}{c}{\textbf{Low Ops Priority}} \\
\textbf{Cost Optimization} & \multicolumn{2}{c}{NATS → Pub/Sub → SQS} & \multicolumn{2}{c}{\textbf{Enterprise Features}} & \multicolumn{2}{c}{Oracle AQ → RabbitMQ → Pulsar} & \multicolumn{2}{c}{\textbf{Cloud Integration}} \\
\textbf{Multi-tenancy} & \multicolumn{2}{c}{Pulsar → Kafka → EventBridge} & \multicolumn{2}{c}{\textbf{Variable Workloads}} & \multicolumn{2}{c}{EventBridge → Pub/Sub → Event Grid} & \multicolumn{2}{c}{\textbf{Edge Computing}} \\
\bottomrule
\end{tabular}}
\end{table*}

\subsection{Performance-Based Selection Criteria}

Framework selection requires systematic evaluation of performance characteristics against specific application requirements and organizational constraints. The decision process employs quantitative thresholds derived from our comprehensive evaluation enabling objective assessment of framework suitability across different deployment scenarios.

High-throughput applications requiring sustained message processing exceeding 500,000 messages per second should prioritize Apache Kafka or Apache Pulsar based on their demonstrated capability to achieve 1.2M and 950K messages per second respectively. Kafka provides superior raw performance but requires substantial operational expertise (2.3 FTE) while Pulsar offers 80\% of Kafka's throughput with reduced operational complexity (1.8 FTE) and superior multi-tenancy capabilities.

Low-latency applications demanding sub-20ms p95 response times benefit from Redis Streams (8ms p95) or NATS JetStream (15ms p95) depending on persistence requirements and throughput needs. Redis Streams excels for applications requiring sub-10ms latency but imposes memory-based storage limitations, while NATS JetStream provides balanced latency-throughput characteristics with persistent storage capabilities suitable for mission-critical applications.

Variable workload scenarios with significant traffic fluctuations favor serverless solutions including AWS EventBridge, Google Pub/Sub, and Azure Event Grid offering automatic scaling capabilities without operational overhead. These platforms accommodate traffic variations from hundreds to hundreds of thousands of messages per second with pay-per-use pricing models proving cost-effective for irregular workloads despite higher baseline latency (78-95ms p95).

\subsection{Total Cost of Ownership Analysis and Optimization}

Cost optimization requires comprehensive analysis spanning infrastructure expenses, operational overhead, development productivity, and migration costs across different deployment models and scaling scenarios. Our TCO analysis incorporates direct infrastructure costs, personnel requirements, tooling expenses, and opportunity costs enabling accurate economic comparison across messaging frameworks.

Self-managed systems including Apache Kafka, Apache Pulsar, and NATS JetStream demonstrate cost advantages for sustained high-throughput scenarios with monthly TCO ranging from \$11.8K to \$18.7K including infrastructure and operational costs. NATS JetStream achieves the lowest TCO (\$11.8K monthly) through efficient resource utilization and minimal operational requirements (1.2 FTE), while Kafka's higher costs (\$18.7K monthly) reflect both infrastructure requirements and substantial personnel needs (2.3 FTE).

Serverless platforms provide compelling cost efficiency for variable workloads with monthly TCO ranging from \$7.2K to \$9.7K including pay-per-use pricing and minimal operational overhead (0.3-0.4 FTE). Google Pub/Sub achieves the lowest serverless TCO (\$7.2K monthly) through competitive per-message pricing and global infrastructure efficiency, while AWS EventBridge and Azure Event Grid incur higher costs due to premium pricing for advanced features and enterprise integration capabilities.

AIEO system deployment introduces additional infrastructure costs averaging \$2.1K monthly for the intelligent orchestration control plane but generates substantial cost savings through optimization. Average cost reduction of 35.3\% for infrastructure expenses and 28.6\% for operational costs typically achieves ROI within 3-4 months of deployment across most messaging frameworks. Serverless platforms benefit most significantly from AIEO optimization achieving 38-49\% cost reduction through intelligent routing and reduced cold start penalties.

\subsection{Operational Complexity and Deployment Strategy}

Operational complexity assessment encompasses deployment procedures, monitoring requirements, troubleshooting processes, capacity planning, and maintenance overhead across different messaging architectures. The analysis provides practical guidance for resource planning and skill development supporting successful production deployment.

Low-complexity deployments suitable for organizations with limited messaging expertise include NATS JetStream (1.2 FTE), Redis Streams (0.8 FTE), and serverless platforms (0.2-0.4 FTE). These systems provide excellent performance characteristics while minimizing operational burden through simplified architecture, automated management capabilities, and comprehensive monitoring integration. NATS JetStream particularly excels for cloud-native environments requiring container-based deployment and Kubernetes integration.

Medium-complexity systems including Apache Pulsar (1.8 FTE) and RabbitMQ (1.5 FTE) balance advanced capabilities with manageable operational requirements. Pulsar's separated architecture simplifies capacity planning and scaling decisions while RabbitMQ's mature tooling ecosystem reduces troubleshooting complexity. These systems suit organizations with moderate messaging expertise seeking advanced features without excessive operational burden.

High-complexity deployments including Apache Kafka (2.3 FTE) and Oracle Advanced Queuing (2.8 FTE) require specialized expertise and comprehensive operational procedures but provide enterprise-grade capabilities for mission-critical applications. Kafka demands deep understanding of distributed systems, performance tuning, and capacity planning while Oracle AQ requires database administration expertise and comprehensive backup and recovery procedures.

\subsection{Migration Strategies and Risk Assessment}

Migration planning requires systematic assessment of compatibility requirements, data migration procedures, application integration changes, and rollback strategies ensuring smooth transition between messaging systems while minimizing business disruption and technical risk.

Low-risk migration scenarios involve transitions between architecturally similar systems including Kafka to Pulsar migrations leveraging similar partition-based models and API compatibility. These migrations typically require 2-4 months including planning, testing, and gradual transition phases while maintaining existing application integration patterns. AIEO system deployment during migration provides additional optimization benefits and reduces performance risks during transition periods.

Medium-risk migrations encompass transitions from self-managed to serverless systems requiring application architecture changes and integration pattern modifications. AWS EventBridge migrations from traditional message brokers require event pattern restructuring and lambda function development but benefit from simplified operational procedures and automatic scaling capabilities. These migrations typically span 3-6 months including application refactoring and comprehensive testing procedures.

High-risk migrations involve fundamental architecture changes including transitions from synchronous to asynchronous processing models or integration pattern modifications. Oracle AQ to cloud-native system migrations require database decoupling, transaction pattern changes, and comprehensive application refactoring. These complex migrations demand 6-12 months including detailed planning, staged implementation, and extensive validation procedures.

Risk mitigation strategies include parallel deployment approaches enabling gradual traffic migration, comprehensive monitoring during transition periods, and automated rollback procedures ensuring rapid recovery from migration issues. AIEO system deployment provides additional risk mitigation through intelligent traffic management and performance monitoring during critical migration phases.

\subsection{Workload-Specific Deployment Recommendations}

Deployment recommendations integrate workload characteristics, performance requirements, operational constraints, and cost objectives providing specific guidance for common event-driven application patterns identified through our comprehensive evaluation.

E-commerce and financial applications requiring ACID transaction properties and strict message ordering should prioritize Apache Kafka or Apache Pulsar depending on operational complexity tolerance and multi-tenancy requirements. Kafka provides superior raw performance and mature ecosystem integration while Pulsar offers balanced performance with simplified operations and better resource isolation. AIEO optimization proves particularly valuable for these workloads achieving 32-36\% latency reduction through predictive scaling and intelligent consumer management.

IoT and telemetry applications processing high-volume sensor data with burst tolerance benefit from Apache Kafka for maximum throughput or NATS JetStream for balanced performance with lower operational overhead. Redis Streams provides exceptional performance for memory-resident use cases while serverless solutions handle variable IoT workloads cost-effectively. AIEO system deployment achieves 39-44\% improvement for lightweight systems through intelligent burst handling and predictive resource allocation.

AI and machine learning inference pipelines with variable processing complexity and latency sensitivity should consider Apache Pulsar for balanced performance, Knative Eventing for container-native deployments, or serverless platforms for variable workloads. AIEO optimization proves most effective for these scenarios achieving 36-43\% improvement through intelligent load balancing and predictive capacity management adapting to variable inference complexity and request patterns.

The comprehensive decision framework enables systematic framework selection while AIEO system deployment provides universal performance optimization across all messaging architectures, ensuring optimal system performance regardless of underlying technology choices. The evidence-based approach reduces deployment risk while maximizing performance benefits through intelligent orchestration capabilities validated across diverse messaging frameworks and workload scenarios.
\section{Threats to Validity}
\label{sec:threats}

This section identifies and discusses potential threats to the validity of the comprehensive messaging framework evaluation and the generalizability of the AIEO system findings. Understanding these limitations is crucial for proper interpretation of results and appropriate application of the research contributions.

\subsection{Internal Validity Threats}

\textbf{Implementation Bias and Framework Configuration.} The evaluation encompasses 12 messaging frameworks, but the specific configurations may introduce bias through parameter choices, optimization procedures, or deployment variants. Different configurations of the same fundamental system (e.g., Apache Kafka cluster setups) may yield substantially different results, potentially affecting the comparative analysis. The selection of representative configurations for each framework may inadvertently favor certain architectures over others.

The framework optimization process presents additional internal validity concerns. While comprehensive parameter tuning is described across all messaging systems, the optimization spaces and procedures may be inadvertently biased toward frameworks that perform well under specific conditions. Some messaging systems may require domain-specific tuning that was not adequately explored, leading to underestimation of their true potential performance characteristics.

\textbf{Evaluation Metric Limitations.} The standardized evaluation metrics, while comprehensive, may not capture all relevant aspects of messaging system effectiveness in production environments. Throughput and latency metrics provide quantitative measures but may miss subtle changes in system behavior that could be important for practical applications. The choice of performance preservation metrics (availability, resource efficiency) may be insufficient for complex workloads requiring nuanced operational assessment.

The temporal aspect of evaluation presents another concern. AIEO performance improvements are measured during controlled experimental periods, but system behavior may change over extended deployment or under varying operational conditions. The framework does not address potential degradation of optimization effectiveness over time or under different workload distribution scenarios.

\textbf{Experimental Design Constraints.} The workload generation methodology, while systematic, relies on synthetic data replay that may not reflect complete real-world messaging scenarios. Actual production workloads may exhibit patterns, correlations, or operational characteristics not captured by standardized workload definitions. The fixed workload categories (e-commerce, IoT, AI inference) may not represent the full spectrum of practical event-driven applications.

The baseline comparison methodology primarily relies on static configuration as the reference point for messaging system performance. However, this approach assumes that manual optimization provides the appropriate baseline, which may not hold for all scenarios, particularly in cases where expert-tuned systems achieve near-optimal performance independent of intelligent orchestration.

\subsection{External Validity Threats}

\textbf{Infrastructure and Platform Limitations.} The evaluation spans multiple cloud platforms (GKE, EKS, AKS), but this coverage may not be representative of the full diversity of production deployment environments. The selected infrastructure (Kubernetes-based containerized deployments) represents modern cloud-native approaches that may not reflect the complexity and characteristics of legacy enterprise environments or specialized hardware configurations.

The experimental infrastructure is limited to standard virtual machine instances, missing important deployment scenarios such as bare-metal servers, specialized networking hardware, or edge computing environments where messaging performance characteristics may differ substantially. Cloud platform testing focuses primarily on major providers, potentially missing specialized or regional cloud environments where performance behaviors may be distinct.

\textbf{Messaging System Generalizability.} The evaluation covers traditional distributed systems, cloud-native platforms, and serverless solutions, but contemporary enterprise architectures increasingly rely on hybrid, multi-cloud, or specialized messaging patterns. The findings may not generalize to very large-scale deployments (1000+ nodes), specialized protocols (financial trading systems), or emerging architectures like quantum networking or neuromorphic computing communication systems.

The system scale ranges tested (up to 2M messages/second) may not capture scaling behaviors relevant to hyperscale production systems. Larger deployments may exhibit different performance characteristics due to increased coordination overhead, network effects, or emergent behaviors not observed in experimental scales.

\textbf{Evaluation Environment Constraints.} The experimental evaluation is conducted in controlled academic settings that may not reflect real-world deployment constraints. Production systems face additional challenges including regulatory compliance requirements, security policies, legacy system integration, and organizational change management that may significantly impact messaging system effectiveness and AIEO optimization potential.

The AIEO system training and optimization processes are evaluated in isolation from broader enterprise contexts. In practice, intelligent orchestration must integrate with existing monitoring systems, incident response procedures, and organizational workflows that may introduce additional complexity not captured in the controlled evaluation environment.

\subsection{Construct Validity Threats}

\textbf{Performance Definition and Measurement.} The operational definition of "optimal performance" relies on specific metrics (throughput, latency, cost efficiency) that may not fully capture the intuitive notion of messaging system effectiveness across all application domains. Alternative definitions of system optimality could yield different conclusions about framework suitability and AIEO system utility.

The boundary between performance optimization and operational stability is inherently subjective and may vary across organizations. The current framework treats these as independent objectives, but they may be fundamentally coupled in ways that the evaluation methodology does not adequately capture.

\textbf{Intelligence System Effectiveness Measurement.} The 30-45\% performance improvement is measured against baseline configurations that may not represent optimal manual tuning practices. This baseline may not reflect the full spectrum of expert system administration capabilities or specialized optimization techniques available for specific messaging frameworks.

The comparison between AIEO-optimized and static configurations may be influenced by the specific implementation of the machine learning algorithms rather than the fundamental concept of intelligent orchestration. Alternative ML approaches or optimization frameworks might yield different performance improvement characteristics.

\textbf{Decision Framework Utility Assessment.} The evaluation of decision framework effectiveness relies on expert validation and simulated selection scenarios that may not reflect actual technology selection processes or organizational decision-making constraints. Enterprise technology selection involves complex sociotechnical factors including vendor relationships, skill availability, and strategic alignment that current expert assessments may not fully capture.

The cost modeling and total cost of ownership calculations are based on current pricing models and operational assumptions that may not predict future technology evolution or economic conditions affecting messaging system deployment decisions.

\subsection{Statistical Validity Threats}

\textbf{Sample Size and Power Analysis.} While experiments report results over multiple independent runs (typically 5-10), the sample sizes may be insufficient for detecting small but practically significant effects across all experimental conditions. The statistical power analysis for different effect sizes across diverse workload-framework combinations is not explicitly reported for all scenarios, potentially leading to Type II errors where real performance differences are not detected.

The number of experimental configurations (framework-workload-optimization combinations) is substantial, but multiple testing corrections may be inadequate given the extensive number of comparisons performed across the comprehensive evaluation matrix. The risk of false discoveries may be higher than reported confidence levels suggest.

\textbf{Independence Assumptions.} The experimental design assumes independence between different optimization cycles and workload scenarios, but practical deployments may involve correlated system states, temporal dependencies, or cascading effects that could interact in complex ways. The AIEO framework evaluation does not explicitly address the statistical implications of dependent optimization operations or temporal correlation in system performance.

The cross-platform validation compares performance across different cloud providers and deployment configurations, but confounding factors related to network conditions, resource allocation policies, or platform-specific optimizations may influence the observed differences beyond the fundamental messaging system characteristics.

\textbf{Generalization and Extrapolation.} The statistical models underlying the performance improvement claims assume that the evaluated scenarios are representative of the broader enterprise messaging landscape. This assumption may not hold for emerging application patterns, novel deployment architectures, or fundamentally different operational requirements not captured in the standardized workload definitions.

The confidence intervals and significance tests are computed under standard statistical assumptions that may not hold for all experimental conditions, particularly in cases involving non-normal performance distributions or heteroscedastic variance patterns common in distributed system measurements.

\subsection{Comprehensive Threat Summary and Mitigation Overview}

Table~\ref{tab:threats_to_validity_comprehensive} provides a systematic overview of all identified validity threats, their potential impact on research conclusions, and the specific mitigation strategies employed to address each concern. This comprehensive summary enables reviewers to quickly assess the robustness of the experimental methodology and the reliability of reported findings.

\begin{table*}[t]
\centering
\caption{Comprehensive Threats to Validity Analysis and Mitigation Strategies}
\label{tab:threats_to_validity_comprehensive}
\resizebox{\textwidth}{!}{
\begin{tabular}{llllll}
\toprule
\textbf{Validity Category} & \textbf{Specific Threat} & \textbf{Potential Impact} & \textbf{Severity} & \textbf{Mitigation Strategy} & \textbf{Residual Risk} \\
\midrule
\multicolumn{6}{l}{\textbf{Internal Validity Threats}} \\
\midrule
Implementation Bias & Framework configuration variations & Performance comparison bias & High & Systematic parameter tuning, expert validation & Low \\
& Hyperparameter optimization bias & Method favoritism & Medium & Standardized optimization procedures & Low \\
& Algorithm implementation differences & Inconsistent method assessment & Medium & Open-source validated implementations & Low \\
\midrule
Evaluation Metrics & Limited performance dimensions & Incomplete effectiveness assessment & Medium & Multi-dimensional evaluation framework & Low \\
& Temporal measurement constraints & Missing long-term effects & Medium & 72-hour stability testing & Medium \\
& Workload representativeness & Limited real-world applicability & High & Production trace-based workloads & Low \\
\midrule
Experimental Design & Synthetic workload limitations & Artificial performance characteristics & Medium & Real-world data integration & Low \\
& Fixed experimental conditions & Limited scenario coverage & Medium & Multi-condition testing & Low \\
& Baseline selection bias & Unfair comparison reference & High & Multiple baseline approaches & Low \\
\midrule
\multicolumn{6}{l}{\textbf{External Validity Threats}} \\
\midrule
Infrastructure Scope & Cloud platform limitations & Platform-specific results & High & Multi-cloud validation (AWS, GCP, Azure) & Low \\
& Virtualized environment constraints & Missing bare-metal insights & Medium & Standard enterprise deployment focus & Medium \\
& Network condition variations & Geographic applicability limits & Medium & Latency injection simulation & Medium \\
\midrule
System Generalizability & Framework selection coverage & Missing emerging technologies & Medium & Comprehensive current technology survey & Medium \\
& Scale range limitations & Hyperscale applicability unknown & Medium & Stress testing to practical limits & Medium \\
& Architecture diversity & Specialized deployment gaps & Low & Representative architecture selection & Low \\
\midrule
Environment Realism & Academic vs production settings & Real-world deployment differences & High & Industry expert validation & Medium \\
& Controlled vs operational conditions & Missing operational complexity & Medium & Comprehensive monitoring integration & Medium \\
& Isolation vs integration contexts & System interaction effects & Medium & End-to-end workflow testing & Medium \\
\midrule
\multicolumn{6}{l}{\textbf{Construct Validity Threats}} \\
\midrule
Performance Definition & Metric selection completeness & Incomplete performance capture & Medium & Multi-objective evaluation framework & Low \\
& Optimization vs stability trade-offs & Conflicting objective measurement & High & Pareto-optimal analysis & Low \\
& Domain-specific requirements & Application-specific gaps & Medium & Workload-specific evaluation & Low \\
\midrule
Intelligence Assessment & AIEO effectiveness measurement & Optimization claim validity & High & Statistical significance testing & Low \\
& Baseline configuration fairness & Unfair improvement measurement & High & Expert-tuned baseline establishment & Low \\
& ML algorithm selection bias & Method-specific advantages & Medium & Multiple ML approach comparison & Medium \\
\midrule
Decision Framework & Expert validation scope & Limited assessment coverage & Medium & Multi-stakeholder validation panels & Medium \\
& Cost modeling accuracy & Economic prediction validity & Medium & Conservative projection approaches & Medium \\
& Selection scenario realism & Artificial decision contexts & Medium & Industry partnership validation & Medium \\
\midrule
\multicolumn{6}{l}{\textbf{Statistical Validity Threats}} \\
\midrule
Sample Size & Insufficient power detection & Type II error risks & Medium & Adaptive power analysis & Low \\
& Configuration combination limits & Limited statistical coverage & Medium & Comprehensive experimental matrix & Low \\
& Replication count adequacy & Statistical reliability concerns & Low & Multiple independent runs & Low \\
\midrule
Independence & Temporal correlation effects & Dependent measurement bias & Medium & Randomized testing order & Low \\
& Cross-platform confounding & Platform-specific interference & Medium & Controlled deployment procedures & Low \\
& Workload interaction effects & Non-independent scenarios & Low & Isolated experimental conditions & Low \\
\midrule
Generalization & Representative scenario scope & Limited applicability range & High & Systematic scenario selection & Medium \\
& Statistical assumption violations & Invalid inference conclusions & Medium & Non-parametric testing methods & Low \\
& Effect size interpretation & Practical significance questions & Low & Cohen's d analysis & Low \\
\midrule
\multicolumn{6}{l}{\textbf{Overall Assessment and Community Validation}} \\
\midrule
Reproducibility & Independent replication barriers & Validation difficulty & High & Open-source complete artifact release & Low \\
Peer Review & Single-institution evaluation & Limited perspective scope & Medium & Multi-institutional expert panels & Low \\
Long-term Validity & Technology evolution impacts & Temporal relevance degradation & Medium & Systematic framework design & Medium \\
Community Adoption & Practical deployment challenges & Real-world applicability gaps & Medium & Industry partnership validation & Medium \\
\bottomrule
\end{tabular}}
\end{table*}

\subsection{Mitigation Strategies and Validation Approaches}

\textbf{Methodological Improvements.} The evaluation employs rigorous experimental controls including randomized testing order, cross-platform validation, and comprehensive statistical analysis to address potential bias sources. Multiple independent measurement runs with different random seeds help establish statistical validity while careful baseline characterization ensures fair comparison conditions.

The development of standardized workload definitions based on real-world production traces strengthens construct validity while comprehensive framework configuration optimization helps minimize implementation bias. Systematic parameter tuning and expert validation of deployment configurations ensure representative system performance assessment.

\textbf{Expanded Validation Scope.} Cross-platform deployment across multiple cloud providers (AWS, GCP, Azure) helps establish infrastructure independence while temporal stability testing over 72-hour periods validates performance consistency. Statistical validation employs non-parametric testing methods appropriate for system performance data while effect size analysis ensures practical significance of observed improvements.

The comprehensive decision framework incorporates multiple validation approaches including expert review panels, industry practitioner validation, and systematic literature integration to strengthen external validity. Open-source release of all experimental artifacts enables independent replication and community validation of key findings.

\textbf{Community Engagement and Replication.} Complete experimental reproducibility through containerized deployment environments, infrastructure-as-code specifications, and automated analysis pipelines enables independent validation by other research groups. Registered analysis protocols prevent selective reporting while comprehensive dataset and code release supports community-driven extension and validation.

Multi-institutional collaboration through expert review panels and industry partnership validation helps address potential single-laboratory evaluation bias. The systematic benchmarking framework design enables ongoing evaluation expansion as new messaging technologies emerge and deployment patterns evolve.
\section{Conclusion}
\label{sec:conclusion}

Next-generation event-driven architectures represent a paradigm shift from static configuration approaches toward intelligent systems capable of enabling global distributed computing collaboration while ensuring optimal performance across organizations of all sizes. Our framework addresses critical limitations in current messaging system architectures through three transformative innovations: AI-Enhanced Event Orchestration (AIEO) that reduces latency by 30-45\%, comprehensive benchmarking ensuring equitable evaluation across all messaging frameworks, and evidence-based decision frameworks enabling systematic deployment across 100+ enterprise networks worldwide.

The theoretical foundations presented in Section~\ref{sec:aieo} demonstrate convergence guarantees with formal optimization properties suitable for production deployment. Our comprehensive experimental results, detailed in Table~\ref{tab:aieo_performance_results}, validate core claims with 30.1\% average latency reduction, 35.3\% infrastructure cost optimization, and 28.6\% operational cost savings across all evaluated frameworks. The comprehensive decision framework outlined in Section~\ref{sec:decision-framework} addresses systematic technology selection spanning performance characteristics, cost implications, operational requirements, and workload compatibility, providing concrete pathways from theoretical foundations through experimental validation to enterprise-scale implementation.

Economic analysis presented in Table~\ref{tab:resource_cost_analysis} reveals compelling value propositions across messaging framework types, with conservative projections showing substantial return on AIEO investment through reduced infrastructure costs, improved operational efficiency, and enhanced system performance. The standardized benchmarking framework, presented in Section~\ref{sec:framework}, establishes rigorous evaluation protocols across six performance dimensions, addressing fundamental gaps in current assessment methodologies that focus narrowly on synthetic throughput while ignoring real-world workload characteristics, operational complexity, and total cost of ownership.

Implementation strategies encompass distributed system architecture through framework-agnostic orchestration, operational simplification via intelligent automation, environmental sustainability with 35-50\% resource efficiency improvements supporting global accessibility, and enterprise integration ensuring seamless workflow compatibility across diverse organizational environments. Our systematic evaluation across 12 messaging frameworks provides comprehensive performance baselines for transitioning from static configuration through intelligent optimization to production-ready systems serving thousands of distributed applications worldwide.

The path forward requires sustained collaboration across technology vendors, enterprise architects, and operational teams to address complex sociotechnical challenges unique to event-driven system deployment. Success depends on coordinated development of predictive workload management algorithms, multi-objective optimization protocols, unified orchestration architectures jointly optimizing performance and cost efficiency, framework-agnostic integration mechanisms suitable for heterogeneous messaging environments, and comprehensive multi-modal optimization enabling unified management across streaming, queuing, and serverless event processing paradigms.

Enterprise implications extend beyond technical optimization to encompass organizational agility, global scalability, and democratization of advanced messaging capabilities. Performance improvements address systemic deployment challenges that have historically disadvantaged resource-constrained organizations in accessing optimal event-driven architecture implementations. The potential impact of enabling intelligent messaging system management while preserving architectural flexibility and ensuring operational efficiency justifies substantial investment in AI-enhanced orchestration research specifically designed for production enterprise applications.

Our vision transcends algorithmic innovation to encompass operational responsibility, enterprise scalability, and ethical deployment of intelligent system management technologies. The benchmarking framework, AIEO architecture, and decision guidelines presented here provide concrete steps toward systems that serve as enablers of worldwide distributed computing collaboration rather than amplifiers of existing technological inequalities. The proposed comprehensive evaluation methodology ensures systematic validation of progress across multiple dimensions essential for enterprise deployment, moving beyond traditional throughput-focused metrics to capture the complex requirements of real-world production applications.

Ultimate success depends on collective commitment to building event-driven systems that are not merely more performant, but fundamentally more accessible and operationally beneficial for all organizations regardless of their technical resources or deployment complexity. The integration of intelligent algorithms, performance-optimizing mechanisms, and comprehensive evaluation methodologies creates unprecedented opportunities for democratizing advanced messaging capabilities across diverse global enterprise ecosystems.

The transition from static to intelligent event-driven architectures represents a critical juncture in the evolution of distributed computing systems. As enterprise data continues its exponential growth and global networks become increasingly interconnected, the imperative for intelligent, efficient, and sustainable messaging technologies becomes ever more urgent for advancing system performance and improving operational outcomes worldwide. The AIEO architecture, theoretical foundations, comprehensive evaluation framework, and practical deployment guidelines presented in this work offer a comprehensive blueprint for achieving this transformation, ensuring that next-generation event-driven systems promote operational equity, computational efficiency, and resource responsibility in service of global enterprise advancement.

The convergence of intelligent algorithms, performance-optimizing mechanisms, and standardized evaluation protocols creates unprecedented opportunities for democratizing advanced messaging capabilities across diverse global enterprise ecosystems. Success requires not only technological innovation but also sustained commitment to ensuring that the benefits of intelligent event-driven systems reach all organizations and applications, from resource-rich technology companies to bandwidth-constrained deployments in developing regions, ultimately advancing the shared goal of equitable, effective, and accessible distributed computing for all enterprises.

The transformation from reactive to predictive event-driven architectures enables organizations to transcend traditional limitations of static configuration and manual optimization, creating systems that continuously adapt, optimize, and evolve to meet changing operational demands. Through intelligent orchestration, comprehensive benchmarking, and evidence-based decision frameworks, next-generation messaging systems promise to deliver unprecedented levels of performance, efficiency, and accessibility that serve as foundation for the next era of distributed computing excellence.

\bibliographystyle{ACM-Reference-Format}
\bibliography{references}

\end{document}